\documentclass[aps,prx,groupedaddress,twocolumn]{revtex4-1}
\usepackage[utf8]{inputenc}
\usepackage{amsmath,bm,amsthm,amsfonts,amssymb}
\usepackage{graphicx}
\usepackage[caption=false]{subfig}
\usepackage{hyperref}
\usepackage{braket}
\usepackage{dsfont}
\usepackage{algorithm, algpseudocode}
\usepackage{algorithmicx}

\newtheorem*{remark}{Remark}

\begin{document}

\title{Graph state representation of the toric code}

\author{Pengcheng Liao}
\affiliation{Institute for Quantum Science and Technology, and Department of
Physics and Astronomy, University of Calgary, Calgary, Alberta, T2N 1N4, Canada }

\author{David L. Feder}
\email[Corresponding author: ]{dfeder@ucalgary.ca}
\affiliation{Institute for Quantum Science and Technology, and Department of
Physics and Astronomy, University of Calgary, Calgary, Alberta, T2N 1N4, Canada }

\begin{abstract}
Given their potential for fault-tolerant operations, topological quantum states 
are currently the focus of intense activity. Of particular interest are 
topological quantum error correction codes, such as the surface and planar 
stabilizer codes that are equivalent to the celebrated toric code. While every 
stabilizer state maps to a graph state under local Clifford operations, the 
graphs associated with topological stabilizer codes remain unknown. We show 
that the toric code graph is composed of only two kinds of subgraphs: star 
graphs (which encode Greenberger-Horne-Zeilinger states) and half graphs. The 
topological order is identified with the existence of multiple star graphs, 
which reveals a connection between the repetition and toric codes. The graph 
structure readily yields a log-depth quantum circuit for state preparation, 
assuming geometrically non-local gates, which can be reduced to a constant 
depth including ancillae and measurements at the cost of increasing the 
circuit width. The results provide a new graph-theoretic framework for the 
investigation of topological order and the development of novel topological 
error correction codes. 
\end{abstract}

\maketitle

\section{Introduction}

Since the idea of topological quantum computation was first introduced by 
Kitaev in the form of the celebrated toric code~\cite{Kitaev1997a, Kitaev1997b,
Kitaev2003}, interest in finding ways to generate topological states and 
implement topological operations has remained strong~\cite{Nayak2008,Stern2013,
Lahtinen2017}, due to the potential for the implementation of fault-tolerant quantum gates with extremely high error thresholds~\cite{Dennis2002}.
In condensed matter physics, topologically ordered states are usually framed as 
the degenerate ground states of a specially chosen gapped local 
Hamiltonian~\cite{Levin2005String,Bombin2006Topological,Xia2009Observation,
Albrecht2016Exponential,Mong2014Universal,Kraus2013Braiding,Sau2010Generic,
Alicea2011NonAbelian}. In the quantum information community, topological models 
are framed in terms of stabilizers, which underpin the framework of quantum 
error correction codes (QECC)~\cite{Gottesman1997}. For example, the toric code 
states are the four-fold degenerate eigenstates of a `Hamiltonian' consisting 
of the the negative sum of $N-2$ toric-code stabilizer generators, where $N$ is 
the number of physical qubits and the degeneracy is connected to the non-zero 
genus of the torus. The code distance of toric code is said to be `macroscopic' 
as it scales with the number of physical qubits. The macroscopic code distance 
is a characteristic of the toric code and is a suitably proxy for the existence 
of topological order~\cite{Bravyi2010Topological}. The toric code is the most 
well-studied topological model, not only because of its apparent simplicity but 
also because all two-dimensional translationally invariant topological 
stabilizer codes (so-called surface or planar codes, depending on the boundary 
conditions) are equivalent to it~\cite{Bombin2012Universal}. 

Every stabilizer state is equivalent to a graph state under local Clifford (LC) 
operations~\cite{Schlingemann2001Stabilizer, Van2004Graphical}. However, little 
is currently known about the structure of the graph states that are 
LC-equivalent to topological stabilizer code states. What is the signature of
topological order in the graph connectivity? What new insights into topology 
and topological QECC might this mapping enable? And, can the graph structure 
point to a specific state preparation and/or logical encoding procedure for 
topological QECC via a quantum circuit? 

In this work, we make the first step of addressing these questions by mapping a 
toric code state to its LC-equivalent graph state, which is denoted as the 
\textit{toric graph state}. The degenerate toric code has two fewer stabilizer 
generators than is the case
for graph states. In order to effect the map, one may supplement the generators
with two closed `string' operators; these consist of contiguous strings of
X and Z gates that encircle the torus, which commute with all toric-code
stabilizer generators and with one other. In this way, one can obtain the graph
state that is LC-equivalent to any of the ground states of the 2D toric code,
depending on the orientation of the string operators; these states are denoted
as \textit{toric graph states} in this work. These kinds of string operators
correspond to the logical X and Z gates in the toric code, and can therefore
map the target toric graph state to any other after the state preparation.
Furthermore, as the strings have length $L\sim\sqrt{N}$, the toric code
distance is macroscopic as it scales with the number of physical qubits. The
macroscopic code distance is a characteristic of the toric code and is a
suitably proxy for the existence of topological order.

We find that the toric graph can be decomposed into 
only two distinct subgraphs: star graphs, where one vertex is connected to all 
other vertices and which define Greenberger-Horne-Zeilinger (GHZ) 
states~\cite{Hein2004} (see for example Fig.~\ref{fig:example_graph}(a)), and half 
graphs~\cite{Erdos1984} (see for example Fig.~\ref{fig:example_graph}(b)). Perhaps 
surprisingly, the macroscopic distance of the toric code is identified with the 
existence of multiple star subgraphs in the toric graph, which reveals a 
connection to the repetition code such as Shor's nine-qubit 
code~\cite{shor1995}.

Despite the fact that the number of edges in the toric
code graph increases as $L^2$, the binary quadratic function defining the graph
adjacency matrix~\cite{Cosentino2009} is shown to be decomposable into a
$\log(N)$ number of operations. Using this insight, we provide explicit quantum 
algorithms to generate toric graph states and to encode arbitrary quantum 
states into the toric QECC in log depth. Furthermore,
because any graph state can be generated in constant depth by including
ancillae that are subsequently projected out via measurements~\cite{hoyer2006},
our algorithm can be expressed in constant depth at the cost of increasing the
circuit width from scaling as $N$ to scaling as $L^3\sim N^{3/2}$. Given the
result in Ref.~\cite{Bombin2012Universal}, this work therefore provides an
algorithm for the preparation of any 2D topological stabilizer code state in
either log depth including only unitary gates or in constant depth allowing for
measurements of ancillae. The mapping to graph states therefore provides an
alternative method to prepare 2D topological stabilizer code states in either 
log depth including only unitary gates or in constant depth allowing for 
measurements of ancillae, complementing currently known 
schemes~\cite{Bravyi2006Lieb, Hamma2008Adiabatic, Aguado2008Entanglement, Konig2009, Huang2015Quantum, Rahmani2020creating}.

This manuscript is organized as follows. The technical background is reviewed 
in Sec.~\ref{sec:background}. The mapping of the toric code stabilizer to the graph is covered in Sec.~\ref{sec:toric_graph_state}, and the resulting graph
structure and its decomposition are discussed. 
Section~\ref{sec:generation_toric_graph_state} covers the construction of the  log-depth quantum circuit that generates a toric graph state and its 
generalization for the preparation of any 2D topological stabilizer code state.
The results are discussed briefly in Sec.~\ref{sec:discussion}. Various 
technical details and proofs are included in the Appendices.

\section{Background and Formalism}
\label{sec:background}

\subsection{Stabilizer states and graph states}
\label{sec:stabilizer_and_graph_state}

Define the Pauli group on $N$ qubits as 
$\mathcal{P}_N:=\{\pm 1,\pm i\}\times\{I,X,Y,Z\}^{\otimes N}$, where
\begin{equation}
X=\begin{pmatrix}
0 & 1\cr 
1 & 0\cr
\end{pmatrix};\quad
Y=\begin{pmatrix}
0 & -i\cr 
i & 0\cr
\end{pmatrix};\quad
Z=\begin{pmatrix}
1 & 0\cr 
0 & -1\cr
\end{pmatrix}\quad
\label{Paulis}
\end{equation}
correspond to the Pauli matrices. The set 
${\mathcal S}:=\{S\in{\mathcal P}_N |\; S\Ket{\psi} = \Ket{\psi}\}$ is said to 
stabilize a state $\ket{\psi}\in\mathcal{H}_2^{\otimes N}$. The set of states 
simultaneously stabilized by $m$ independent operators $\{S_1,\ldots,S_m\}$ from 
$\mathcal{P}_N$ then generate a state subspace of dimension $2^{N-m}$. When 
$m=N$, the subspace contains only one state called the stabilizer state, and 
the $N$ independent operators are the generators of ${\mathcal S}$.

Graph states are special stabilizer states where the stabilizer generators are 
related to simple graphs~\cite{Hein2004}. Given a graph $G=(V,E)$, where 
$|V|=N$, the corresponding graph state is
\begin{align}
\label{eq:graph_state_CZij}
    \Ket{G} =  \prod_{(i,j)\in E \atop i<j} \operatorname{CZ}(i,j) H^{\otimes N} 
\Ket{0^N},
\end{align}
for which the stabilizer generators are
\begin{equation}
S_{i} = X_{i}  \prod_{(i,j)\in E } Z_{j}.
\end{equation}
A graph can be represented by its adjacency matrix $A\in\mathbb{Z}_2^{N\times N}$, where $A_{ij}=1$ iff $(i,j)$ is an edge in $E$.
With adjacency matrix $A$, the graph state can also be written in terms of its binary quadratic form~\cite{Cosentino2009}
\begin{equation}
\Ket{G} = \frac{1}{\sqrt{2^N}}\sum_{q\in\set{0,1}^N} (-1)^{f_G(q)}\Ket{q},
\label{eq:graph_state_explicit_form}
\end{equation}
in which $f_G(q)=\sum_{i<j}A_{ij} q_i q_j \mod 2$ is a quadratic Boolean 
function. Unless stated otherwise, the addition of binary variables is 
performed $\mod 2$. There is therefore a useful correspondence among simple 
graphs, graph states, and quadratic Boolean functions. 

\begin{figure}[t]
\includegraphics[width=0.5\textwidth]{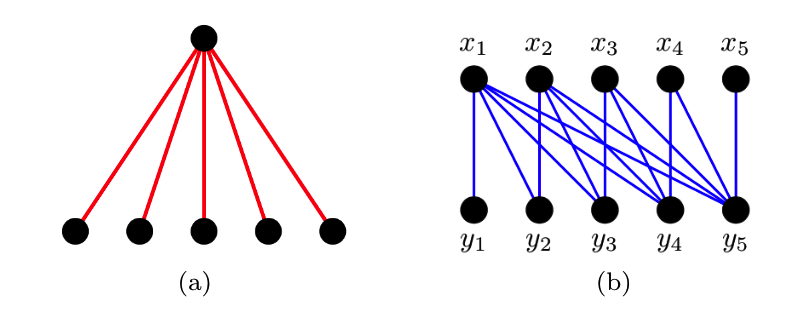}
\caption{(Color online) The two graphs comprising the toric graph: (a) star graph, (b) half graph.}
\label{fig:example_graph}
\end{figure}

As discussed in Sec.~\ref{sec:toric_graph_state}, the toric code maps to a 
graph, called the toric graph in what follows, that can be decomposed into two 
types of subgraphs: star and half graphs. These are reviewed here, and examples
are shown in Fig.~\ref{fig:example_graph}. The star graph on $m$ vertices is 
the complete bipartite graph $K_{m-1,1}$, as shown in Fig.~\ref{fig:example_graph}(a). 
Because the star graph is LC-equivalent to the complete graph 
$K_m$~\cite{Hein2004}, the label of the large-degree vertex is arbitrary. 
Without loss of generality, the non-zero elements of the adjacency matrix are 
then $A_{mi}=A_{im}=1$ for $i=1,\ldots,m-1$, or alternatively
\begin{equation}
\label{eq:A^star_ij}
A_{ij}^{(\rm star)}=\delta_{i,m}\theta_{j,m-1}+\delta_{j,m}\theta_{i,m-1},
i,j\in\{1,\ldots,m\},
\end{equation}
where $\delta_{i,j}$ and $\theta_{i,j}$ are the usual Kronecker and Heaviside
theta functions, respectively:
\begin{equation}
\delta_{i,j}=\begin{cases}
1 & j=i\\
0 & {\rm otherwise}
\end{cases}\;{\rm and}\;
\theta_{i,j}=\begin{cases}
1 & i\leq j\\
0 & {\rm otherwise}.
\end{cases}
\end{equation}
The star graph state can then be written as
\begin{eqnarray}
\Ket{G_{\text{star}}} 
&=&\frac{\Ket{+}^{\otimes m-1}\Ket{0} +\Ket{-}^{\otimes m-1}\Ket{1}}{\sqrt{2}}
\nonumber \\
&=& \frac{1}{\sqrt{2^m}}\sum_{q\in\set{0,1}^{m}} (-1)^{f_{\rm star}(q)}\Ket{q},
\end{eqnarray}
where $f_{\rm star}(q) =(q_1+\cdots+ q_{m-1})\cdot q_m
=q_m{\mbox P}({\rm q}_{m-1})$; here P is the parity operator acting on the 
length $m-1$ bit string ${\rm q}_{m-1}\equiv q_1\oplus\cdots\oplus q_{m-1}$. 
Evidently, $\Ket{G_{\text{star}}}$ is locally equivalent to an $m$-qubit GHZ 
state
\begin{equation}
\label{eq:lcequiv_GHZ_star}
I^{\otimes m-1}\otimes H\Ket{G_{\text{star}}}
=\frac{\ket{+}^{\otimes m}+\ket{-}^{\otimes m}}{\sqrt{2}}.
\end{equation}

The second example is the half graph $K_{n,n}^{(\rm half)}$ (our notation), a 
$2n$-vertex balanced bipartite graph where the $n$ vertices $x_i$ 
($i=1,\ldots,n$) in one bipartition share an edge with the $n$ vertices $y_j$ 
($j=1,\ldots,n$) in the other whenever $i\le j$. An example is shown in 
Fig.~\ref{fig:example_graph}(b). Alternatively, the non-zero entries of the adjacency 
matrix can be written as
\begin{equation}
\label{eq:half_A_xiyj}
A_{x_i y_j}^{(\rm half)}=\theta_{i,j},\quad i\in\{1,\ldots,n\},\;
j\in\{1,\ldots,n\}.
\end{equation}
The corresponding graph state can be written as
\begin{equation}
\Ket{K_{n,n}^{(\rm half)}} = \frac{1}{\sqrt{2^{2n}}}\sum_{{q,p\in\set{0,1}^n}}(-1)^{f_{\rm half}(q,p)}\Ket{q,p},
\label{eq:half_state_polar_form}
\end{equation}
where the associated quadratic function is 
$f_{\rm half}(q,p) = \sum_{i\le j}q_ip_j$.

\begin{figure}[t]
\includegraphics[width = 0.6\columnwidth]{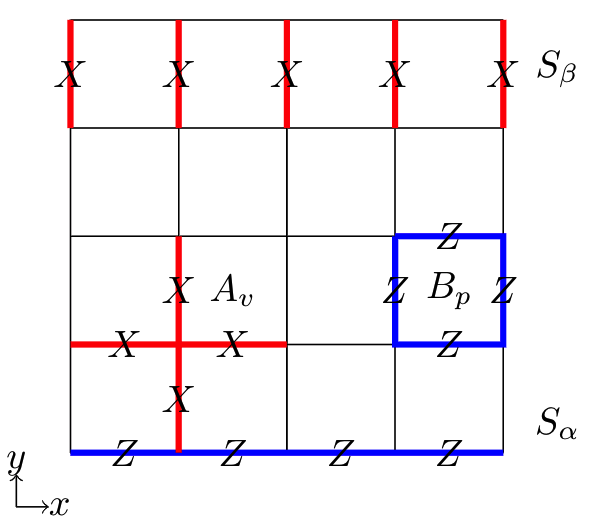}
\caption{(Color online) Representation of operators in the 2D toric code. 
Qubits reside on the
edges of the lattice (not shown), on which $X$ and $Z$ gates are applied (red
and blue lines, respectively). The red plus and blue square represent the star 
$A_v$ and plaquette $B_p$ operator, respectively; the corresponding string 
operators $S_{\beta}$ and $S_{\alpha}$ are also depicted.}
\label{fig:2d_toric_lattice} 
\end{figure}

One can directly use Eq.~(\ref{eq:graph_state_CZij}) to construct the quantum 
circuit to generate an arbitrary graph state $\Ket{G}$. As the CZ gates share 
no common qubits, they can all be implemented in parallel, yielding a 
$\Delta(G)$-depth circuit~\cite{hoyer2006}, where $\Delta(G)$ is the (maximum)
vertex degree of the graph. For the $n$-vertex star graph and $2n$-vertex half 
graph discussion above, preparing the corresponding graph states then seems to
require a $O(n)$-depth quantum circuit. In fact, as we show below, these two 
special graph states can be generated with a log-depth quantum circuit.

\subsection{2D toric code}
\label{sec:2d_toric_code}

The toric code on a square lattice is defined on an $L\times L$ grid with 
periodic boundary conditions in both directions, so that the system geometry 
corresponds to a two-torus. A physical qubit is placed on every edge, so that 
there are a total of $N=2L^2$ qubits. The stabilizer 
generators are
\begin{equation}
    A_v= \prod_{i\in v} \sigma^x_i,\quad
    B_p= \prod_{i\in p} \sigma^z_j,
\end{equation}
where the first sum is over the nearest qubits surrounding a given vertex $v$
of the lattice, corresponding to a `star' operator, while the second is over 
the nearest qubits surrounding the center $p$ of a square, corresponding to a
`plaquette' operator. These operators are depicted in 
Fig~\ref{fig:2d_toric_lattice}. 

The star and plaquette operators can share at most two edges, so that the $A_v$ 
and $B_p$ commute. However, they are not entirely independent because
\begin{equation}
\prod_{v} A_v  = I,\quad \prod_{p} B_p  = I,
\label{eq:dependence_toric_generators}
\end{equation}
where $I$ is the identity. There are $N-2$ independent stabilizer generators 
and the ground subspace of the associated Hamiltonian
\begin{equation}
\label{eq:toric_hamiltoian}
H =  -\sum_{v} A_v  -\sum_{p} B_p
\end{equation}
is  four-fold degenerate. In order to specify one state in the degenerate 
subspace, we add two more stabilizer generators $S_{\alpha}$ and $S_{\beta}$. 
One choice for the $S_{\alpha}$ ($S_{\beta}$) corresponds to a string of $Z$ 
($X$) gates applied to the qubits residing only on vertical (horizontal) edges 
of a given row of the lattice. The choice of row is unimportant because 
of the translational invariance of the system; also, the rows for 
$S_{\alpha}$ and $S_{\beta}$ can coincide because the operations are on a 
different set of qubits.

\subsection{Local Clifford equivalence and the symplectic representation}
\label{sec:symplectic}

In this work, we make heavy use of the symplectic representation of Pauli 
operators~\cite{Gottesman1997}. As this formalism is not widely employed, we 
briefly review the notation here, closely following the notation in
Ref.~\cite{Van2004Graphical}. Neglecting overall phases, a single Pauli 
matrix~(\ref{Paulis}) can be written as $\sigma=Z^uX^v$, where $u,v\in\{0,1\}$.
Alternatively, they can be represented as a binary tuple, $I\to(0|0)$, 
$X\to(0|1)$, $Z\to(1|0)$, and $Y\to(1|1)$ or equivalently the vectors
\begin{equation}
I=\begin{pmatrix}
0\cr
0\cr
\end{pmatrix};\quad
X=\begin{pmatrix}
0\cr
1\cr
\end{pmatrix};\quad
Y=\begin{pmatrix}
1\cr
1\cr
\end{pmatrix};\quad
Z=\begin{pmatrix}
1\cr
0\cr
\end{pmatrix}.
\label{Paulis2}
\end{equation}

The generalization to $N$ qubits 
$\sigma_1\otimes\cdots\otimes\sigma_N$ is then $(u_1\ldots u_N|v_1\ldots v_N)
\in\mathbb{Z}_2^{2N}$, i.e.,\ a $2N$-dimensional binary vector. For example, 
\begin{equation}
X\otimes Z=(01|10)=\begin{pmatrix}
0 \cr
1 \cr
1\cr
0\cr
\end{pmatrix}.
\end{equation}
The $N$ stabilizer generators that uniquely define an $N$-qubit state can then 
expressed as full-rank $2N\times N$-dimensional matrix $S$; for example, the 
graph state corresponding to the two-vertex path graph $P_2$ is defined by the 
stabilizer generators $S_1=X\otimes Z$ and $S_2=Z\otimes X$, which are combined 
in the symplectic notation:
\begin{equation}
S_{P_2}=\begin{pmatrix}
0 & 1\cr
1 & 0\cr
1 & 0\cr
0 & 1\cr
\end{pmatrix},
\end{equation}
In general, the sympletic notation of the stabilizer generators for every graph 
state $\Ket{G}$ is 
\begin{equation}
\label{eq:A|I}
S=
\begin{pmatrix}
A\cr
I\cr
\end{pmatrix},    
\end{equation} 
where $A$ is the adjacency matrix of graph $G$. This form is referred as the 
standard form of the stabilizer for graph states in this work.
The central advantage of this formulation is that, because all of the 
stabilizer generators are mutually commuting, the matrix $S$ is automatically 
self-orthogonal under the symplectic inner product $S^{T}JS=0$, where 
\begin{equation}
J=\begin{pmatrix}
{\bf 0} & I\cr
I & {\bf 0}\cr
\end{pmatrix}
\end{equation}
is the symplectic metric tensor (note that the sign convention is not the same 
as the symplectic algebra in classical Hamiltonian mechanics), and ${\bf 0}$ 
represents the all-zero matrix. 

Clifford operations, which maps the Pauli group to itself under conjugation, 
then correspond to $2N\times 2N$ matrices $Q$ that preserve the metric, i.e.\ 
$Q^TJQ=J$~\cite{dehaene2003}. Local Clifford gates refer to those which are 
tensor products of local gates acting on single qubits. Two quantum states are 
local Clifford equivalent if they can be mapped to each other by local Clifford 
operations. Clifford operations that transform a stabilizer generator matrix 
$S$ to another $S'$ can always be written in the form $S'=QSR$, where $R$ is an 
$N\times N$ invertible matrix corresponding to a basis 
change~\cite{Van2004Graphical}. Restricting to local Clifford gates further 
implies that $Q$ can be partitioned into four $N\times N$ blocks, each of which 
is diagonal. For example, if we partition all qubits into two complementary 
sets $R_1\cup R_2=\set{1,\ldots,N}$, then
\begin{equation}
H_{R_2}= \bigotimes_{i\in R_1}I\bigotimes_{j\in R_2}H_j
\end{equation}
is a local Clifford operation and its sympletic representation is
\begin{equation}
Q = 
\begin{pmatrix}
I_{R_1} & \bm{0} & \bm{0} & \bm{0} \\
\bm{0}  & \bm{0} & \bm{0} & I_{R_2}\\
\bm{0} & \bm{0}  & I_{R_1}& \bm{0}\\
\bm{0} & I_{R_2} & \bm{0} & \bm{0}
\end{pmatrix}.
\label{eq:hadamard_conjugation}
\end{equation}
The transformation of the stabilizer is then effected by ordinary matrix-vector 
multiplication. These results imply a specific procedure to map any 
stabilizer generator matrix to standard (graph) form, which is used in this 
work to derive the (non-unique) graph that stabilizes the toric code.

\section{Toric graph state}
\label{sec:toric_graph_state}

In this section, the toric code is first expressed in the symplectic notation.
Then the toric graph state is obtained for one of the toric code states by
specifying the two string operators. Finally, the toric graph is shown to be
decomposable into star and half graphs, and prove that the subgraphs consisting 
of multiple star graphs contribute to the macroscopic distance of the toric 
QECC.  

\subsection{Symplectic representation of the toric code }
\label{subsec:symplectic_toric}

In the toric code, qubits are located on the edges of a regular $L\times L$
square lattice thus there are $N=2L^2$ qubits in total. It is convenient to distinguish the qubits situated on 
horizontal ($x$) and vertical ($y$) edges, whose locations on the grid are 
denoted by $(i,j,x)$ and $(i,j,y)$, respectively, where $i,j\in\{1,L\}$.
The star operator $A_v^{ij}$ centered at coordinate $(i,j)$ then includes
qubits with labels $\{(i-1,j,x), (i,j,x), (i,j-1,y), (i,j,y)\}\mod L$;
likewise, the plaquette operator $B_p^{ij}$ with the coordinate $(i,j)$ located 
at the bottom left of a given plaquette includes qubits with labels $\{(i,j,x),
(i,j+1,x), (i,j,y), (i+1,j,y)\}\mod L$. Given that the star and plaquette terms 
apply $X$ and $Z$ gates, respectively, they can be expressed in the symplectic 
representation as
\begin{equation}
A^{ij}_{\rm bin} = \begin{pmatrix}
0\\v^{ij}
\end{pmatrix};\quad
B^{ij}_{\rm bin} = 
\begin{pmatrix}
p^{ij}\\0
\end{pmatrix},
\end{equation}
where $A^{ij}_{\rm bin},B^{ij}_{\rm bin}$ are $4L^2$-length vectors, and
$v^{ij},p^{ij}\in\mathbb{Z}_2^{2L^2}$ are binary strings with elements defined 
by
\begin{align}
v^{ij}_{lmd}=\delta_{i,l}\delta_{j,m}+\delta_{i-1,l}\delta_{j,m}\delta_{d,x}
+\delta_{i,l}\delta_{j-1,m}\delta_{d,y};
\nonumber \\
p^{ij}_{lmd}=\delta_{i,l}\delta_{j,m}+\delta_{i,l}\delta_{j+1,m}\delta_{d,x}
+\delta_{i+1,l}\delta_{j,m}\delta_{d,y},
\label{eq:vp}
\end{align}
where $l,m\in\{1,\ldots,L\}$ and $d\in\set{x,y}$. The toric code stabilizer, without
the string operators, therefore consists of the antidiagonal block matrix
\begin{equation}
\label{eq:bina_repre_star_plaque}
S =
\begin{pmatrix}
{\bf 0} & Z_p \\
X_v& {\bf 0}
\end{pmatrix},
\end{equation}
where the columns of the $2L^2\times L^2$ matrices $X_v$ and $Z_p$ correspond 
respectively to the $v^{ij}$ and $p^{ij}$, $i,j\in\{1,L\}$.

With the expressions~(\ref{eq:vp}), it is straightforward to prove the 
relations~(\ref{eq:dependence_toric_generators}) within the symplectic 
notation:
\begin{equation}
\prod_{v} A_v=\prod_v\begin{pmatrix}
{\bf 0}\\
X_v\\
\end{pmatrix}
=\begin{pmatrix}
{\bf 0}\\
\sum_{ij}v^{ij}\\
\end{pmatrix}.
\end{equation}
Note that evaluation of the terms in the sum above (and in what follows) is 
accomplished via bitwise exclusive or (XOR), with $0+0=1+1=0$ and $0+1=1+0=1$.
The resulting bitstring is then
\begin{eqnarray}
\sum_{ij}v^{ij}&=&\sum_{ij}\bigoplus_{lmd}v^{ij}_{lmd}\nonumber \\
&=&\bigoplus_{lmd}\sum_{ij}\left(\delta_{i,l}\delta_{j,m}
+\delta_{i-1,l}\delta_{j,m}\delta_{d,x}+\delta_{i,l}\delta_{j-1,m}\delta_{d,y}
\right)\nonumber \\
&=&\bigoplus_{lm}\left(\bigoplus_d1+\bigoplus_{d}\delta_{d,x}+\bigoplus_{d}\delta_{d,y}\right)
={\bf 0},
\end{eqnarray}
which yields 
\begin{equation}
\prod_vA_v=\begin{pmatrix}
{\bf 0}\\
{\bf 0}\\
\end{pmatrix}=I,
\end{equation}
as expected. The second condition in Eq.~(\ref{eq:dependence_toric_generators})
is found analogously. Note that in this work, the $\oplus$ notation represents
a direct sum rather than an XOR operation.


\subsection{Star operators}
\label{sec:star_operators}

To bring the stabilizer~(\ref{eq:bina_repre_star_plaque}) into the standard 
form as in Eq.~(\ref{eq:A|I}), the $2L^2\times 2L^2$ submatrix $(0|X_v)$ must be transformed into an 
identity. Due to the constraint~(\ref{eq:dependence_toric_generators}), the 
rank of $X_v$ is only $L^2-1$. The first step in the procedure is to form a 
full-rank $L^2-1$-dimensional matrix by taking linear combinations of the $X_v$ 
vectors to obtain a zero column vector. This column is replaced by $S_\beta$, 
and finally column permutations yield an $L^2\times{L^2}$ identity submatrix.
The remaining $L^2\times{L^2}$ identity submatrix will be obtained from the 
plaquette operators in Sec.~\ref{sec:plaquette_operators}.

The first linear combination is achieved by multiplying $X_v$ on the right by
$R^xT^x$, where $T^x,R^x\in\mathbb{Z}_2^{L^2\times L^2}$ are invertible 
matrices with elements defined by
\begin{align}
R^x_{kn,ij}&=\theta_{i,k}\delta_{j,n};\nonumber \\
T^x_{kn,ij}&=\begin{cases}
\delta_{i,k}\theta_{j,n} & i=1\\
\delta_{i,k}\delta_{j,n} & {\rm otherwise.} \\
\end{cases}
\end{align}

Consider first the action of $R^x$: it transforms the star operator represented 
by $v^{ij}$ to ${v^{ij}}^\prime$.
The column vectors of $X_vR^x$ are
\begin{eqnarray}
{v^{ij}}^{\prime}&=&\sum_{kn}v^{kn}R^x_{kn,ij}=\sum_{kn} v^{kn}\theta_{i,k}
\delta_{j,n}=\sum_{k=i}^{L}v^{kj}\nonumber \\
&=&\sum_{k=i}^{L}\bigoplus_{lmd}\left(\delta_{k,l}\delta_{j,m}
+\delta_{k-1,l}\delta_{j,m}\delta_{d,x}+\delta_{k,l}\delta_{j-1,m}\delta_{d,y}
\right)\nonumber \\
&=&\sum_{k=i}^{L}\bigoplus_{lmd} \delta_{d,x}\delta_{j,m}\left(\delta_{k,l}+\delta_{k-1,l}\right)\nonumber \\
&+&\sum_{k=i}^{L}\bigoplus_{lmd}\delta_{d,y}\delta_{k,l}\left(\delta_{j,m}
+\delta_{j-1,m}\right).
\end{eqnarray}
Let's evaluate the first term above:
\begin{eqnarray}
&&\bigoplus_{lmd}\sum_{k=i}^{L}\delta_{d,x}\delta_{j,m}\left(\delta_{k,l}
+\delta_{k-1,l}\right)\nonumber \\
&&\qquad=\bigoplus_{lmd}\delta_{d,x}\delta_{j,m}\Big(\delta_{i,l}
+\delta_{i-1,l}+\delta_{i+1,l}+\delta_{i,l}\nonumber \\
&&\qquad +\ldots+\delta_{L-1,l}+\delta_{L-2,l}
+\delta_{L,l}+\delta_{L-1,l}\Big)\nonumber \\
&&\qquad =\bigoplus_{lmd}\delta_{d,x}\delta_{j,m}\left(\delta_{i-1,l}+\delta_{L,l}\right),
\end{eqnarray}
where only the endpoints in the sum are unpaired and therefore remain.
One then obtains the matrix elements
\begin{eqnarray}
{v^{ij}_{lmd}}^{\prime}
&=& \left(\delta_{L,l}+\delta_{i-1,l}\right)\delta_{d,x}\delta_{j,m}\nonumber \\
&+&\theta_{i,l}\left(\delta_{j,m}+\delta_{j-1,m}\right)\delta_{d,y}.
\label{eq:vijprime}
\end{eqnarray}

Next consider the action of $T^x$: 
\begin{eqnarray}
{v^{ij}}^{\prime\prime}&=&\sum_{kn}{v^{kn}}^{\prime}T^x_{kn,ij}
\nonumber \\
&=&\sum_{kn}{v^{kn}}^\prime\delta_{i,k}\times\begin{cases}
\theta_{j,n} & i=1\\
\delta_{j,n} & {\rm otherwise}
\end{cases}
\end{eqnarray}
which transform ${v^{ij}}^{\prime}$ to ${v^{ij}}^{\prime\prime}$.
When $i\neq 1$, one has ${v^{ij}}^{\prime\prime}={v^{ij}}^{\prime}$, but for $i=1$ one obtains
\begin{equation}
{v^{1j}}^{\prime\prime}
=\sum_{n=j}^L{v^{1n}}^{\prime}.
\end{equation}
Using Eq.~(\ref{eq:vijprime}) and the fact that $i-1=L$ when $i=1$ due to the
periodic boundary conditions, the matrix elements become
\begin{eqnarray}
\label{eq:v1jprimeprime}
{v^{1j}_{lmd}}^{\prime\prime}&=&\sum_{n=j}^L\delta_{d,y}\left(\delta_{n,m}
+\delta_{n-1,m}\right)\theta_{1,l}\nonumber \\
&=&\delta_{d,y}\left(\delta_{L,m}+\delta_{j-1,m}\right).
\end{eqnarray}
Again because of the periodic boundary conditions, ${v^{11}}^{\prime\prime}
={\bf 0}$, which shows explicitly that the rank of $X_vR^xT^x$ is reduced by 
one. In order to make $X_v$ a full-rank matrix, the zero column is replaced by 
the string operator $S_\beta=({\bf 0}|s_\beta)$, where 
${s_\beta} =\bigoplus_{lmd} \delta_{d,y}\delta_{L,m}$. One then obtains
\begin{equation}
\label{eq:v1jppprime}
v^{1j\prime\prime\prime}=v^{1j+1\prime\prime}+s_\beta
=\bigoplus_{lmd}\delta_{d,y}\delta_{j,m}.
\end{equation}
All other column vectors remain unchanged: $v^{ij\prime\prime\prime}
=v^{ij\prime\prime}=v^{ij\prime}$ when $i\ne 1$.

It remains to show that one can extract an $L^2\times L^2$ identity submatrix 
from $X_v^{\prime\prime\prime}$, whose columns are the 
${v^{ij}}^{\prime\prime\prime}$. Consider two complementary subsets of row
indices $R_1$ and $R_2$, defined as
\begin{eqnarray}
R_1&:=&\{1,\ldots,L-1\}_x\times\{1,\ldots,L\}_x\nonumber\\
&\cup&\set{1}_y\times\set{1,\ldots,L}_y; \nonumber \\
R_2&:=&\set{L}_x\times \set{1,\ldots,L}_x\nonumber\\
&\cup&\set{2,\ldots,L}_y\times\set{1,\dots,L}_y,
\label{eq:row_index_xz}
\end{eqnarray}
where $\{\cdots\}_d\times\{\cdots\}_d$ represents the row and column indices of
the original square lattice and $d$ indicates a horizontal $x$ or vertical $y$
qubit. Clearly, both $R_1$ and $R_2$ contain $L^2$ rows. Combining the results
of Eqs.~(\ref{eq:vijprime}) and (\ref{eq:v1jppprime}), one obtains the elements
\begin{eqnarray}
{v^{ij}_{lmd}}_{R_1}^{\prime\prime\prime}&=&\delta_{j,m}\times
\begin{cases}
\delta_{i,l}\delta_{d,y} & i=1\\
\delta_{i-1,l}\delta_{d,x} & i\geq 2;\\
\end{cases}
\nonumber \\
{v^{ij}_{lmd} }_{R_2}^{\prime\prime\prime}&=&
\begin{cases}
\theta_{2,l}\delta_{j,m}\delta_{d,y} & i=1\\
\delta_{L,l}\delta_{j,m}\delta_{d,x}
+\theta_{i,l}\left(\delta_{j,m}+\delta_{j-1,m}\right)\delta_{d,y} & i\geq 2.\\
\end{cases}
\nonumber \\
\label{eq:vprimeprimeprime}
\end{eqnarray}
There is only one non-zero element in each bitstring
${v^{ij}_{R_1}}^{\prime\prime\prime}$ and its location is unique. Therefore, 
${X_v}_{R_1}^{\prime\prime\prime}$ is an $L^2\times L^2$ identity matrix after 
appropriate permutation of columns, which will be effected in 
Sec.~\ref{sec:hadamard_conjugation}.

\subsection{Plaquette operators}
\label{sec:plaquette_operators}

The plaquette operators are treated in much the same way as the star operators
discussed in Sec.~\ref{sec:star_operators}: form linear combinations of the 
$Z_p$ column vectors to obtain a zero column vector, then replace this with the
string operator $S_\alpha$ to make $Z_p$ a full-rank matrix, and finally use 
linear combinations again to extract an $L^2\times L^2$ submatrix.

Define invertible operators $T^z,R^z\in\mathbb{Z}_2^{L^2\times L^2}$ with 
elements
\begin{align}
R^z_{kn,ij} =&\theta_{k,i}\delta_{j,n};\\
T^z_{kn,ij} =& \delta_{i,k}
\begin{cases}
\theta_{n,j} &i=L\\
\delta_{n,j} &{\rm otherwise.}
\end{cases}
\end{align}
The column vector of $Z_pR^z$ is
\begin{equation}
\begin{aligned}
p^{ij\prime}
=& \sum_{kn}  p^{kn} R^z_{kn,ij} 
= \sum_{k} p^{kj}\theta_{k,i} 
=\sum_{k=1}^{i}p^{kj}\\
=& \sum_{k=1}^{i}\bigoplus_{lmd}  (\delta_{k,l}\delta_{j,m}
+\delta_{k,l}\delta_{j+1,m}\delta_{d,x}
+\delta_{k+1,l}\delta_{j,m}\delta_{d,y}) \\
=&\bigoplus_{lmd}\delta_{d,x}(\delta_{j,m}+\delta_{j+1,m}) \sum_{k=1}^{i}\delta_{k,l}\\
+&\bigoplus_{lmd}\delta_{d,y}\delta_{j,m}\sum_{k=1}^{i}(\delta_{k,l}+\delta_{k+1,l}) \\
=&\bigoplus_{lmd}\theta_{l,i}(\delta_{j,m}+\delta_{j+1,m})\delta_{d,x}
+(\delta_{1,l}+\delta_{i+1,l})\delta_{j,m}\delta_{d,y}.
\end{aligned}    
\end{equation}
The column vector of $Z_pR^zT^z$ is ${p^{ij}}^{\prime\prime}={p^{ij}}^{\prime}$ 
when $i\ne L$. When $i=L$, one can make use of the periodic boundary conditions
to obtain
\begin{eqnarray}
p^{Lj\prime\prime}&=& \sum_{kn} p^{kn\prime} T^z_{kn,Lj}
= \sum_{n} p^{Ln\prime}\theta_{n,j}
= \sum_{n=1}^{j}p^{Ln\prime}\nonumber \\
&=&\bigoplus_{lmd}\sum_{n=1}^{j}(\delta_{n,m}+\delta_{n+1,m})\theta_{l,L}\delta_{d,x}\nonumber \\
&=&\bigoplus_{lmd}(\delta_{1,m}+\delta_{j+1,m})\delta_{d,x}.
\end{eqnarray}

Again, one obtains a zero vector, ${p^{LL}}^{\prime\prime}=\bm{0}$. In order to 
yield a full-rank matrix for $Z_p$, we include the string operator 
$S_\alpha=\left(s_\alpha|{\bf 0}\right)$, where 
$s_\alpha=\bigoplus_{lmd} \delta_{d,x}\delta_{m,1}$. Then
\begin{equation}
{p^{Lj}}^{\prime\prime\prime} = {p^{L,j-1}}^{\prime\prime}+s_\alpha= \bigoplus_{lmd}\delta_{j,m}\delta_{d,x}. 
\end{equation}
All other column vectors remain unchanged: ${p^{ij}}^{\prime\prime\prime}={p^{ij}}^{\prime\prime}={p^{ij}}^{\prime}$ when $i\ne L$.

The column vector of matrix ${Z_p}^{\prime\prime\prime}$ is 
${p^{ij}}^{\prime\prime\prime}$, from which one can extract another
$L^2\times L^2$ identity submatrix. Again using the complementary subsets of
row indices $R_1$ and $R_2$, Eq.~(\ref{eq:row_index_xz}), the elements of the 
plaquette bitstrings become
\begin{eqnarray}
{p^{ij}_{lmd}}_{R_1}^{\prime\prime\prime}&=&\begin{cases}
\theta_{l,i}(\delta_{j,m}+\delta_{j+1,m})\delta_{d,x}
+\delta_{1,l}\delta_{j,m}\delta_{d,y} & i<L\\
\theta_{l,L-1}\delta_{j,m}\delta_{d,x} & i=L;
\end{cases}\nonumber \\
{p^{ij}_{lmd}}_{R_2}^{\prime\prime\prime}&=&\delta_{j,m}\times\begin{cases}
\delta_{i+1,l}\delta_{d,y} & i<L\\
\delta_{l,L}\delta_{d,x} & i=L.
\end{cases}
\label{eq:pprimeprimeprime}
\end{eqnarray}
As was the case for the star operators, there is only one non-zero element in 
each bitstring ${p^{ij}_{R_2}}^{\prime\prime\prime}$ and its location is 
unique. Therefore, ${Z_p}_{R_2}^{\prime\prime\prime}$ is another 
$L^2\times L^2$ identity matrix after appropriate permutation of columns, 
which will be effected in Sec.~\ref{sec:hadamard_conjugation}. Combining these 
results with those in Sec.~\ref{sec:star_operators}), the toric code stabilizer 
in sympletic form, Eq.~(\ref{eq:bina_repre_star_plaque}), is now transformed to
\begin{equation}
\label{eq:binary_matrix_after_basis_change}
S\to\begin{pmatrix}
\bm{0}   & {Z_p}^{\prime\prime\prime}\\
{X_v}^{\prime\prime\prime} & \bm{0}
\end{pmatrix}=
\begin{pmatrix}
\bm{0}   & {Z_p}^{\prime\prime\prime}_{R_1}\\
\bm{0}   & {Z_p}^{\prime\prime\prime}_{R_2}\\
{X_v}_{R_1}^{\prime\prime\prime} & \bm{0} \\
{X_v}_{R_2}^{\prime \prime\prime}& \bm{0}\\
\end{pmatrix}.
\end{equation}

\subsection{Transformation to standard form}
\label{sec:hadamard_conjugation}

To convert the stabilizer~(\ref{eq:binary_matrix_after_basis_change}) to
standard form, one first applies Hadamard operations to the $R_2$ qubits,
\begin{equation}
\label{eq:QS=AB}
Q\begin{pmatrix}
\bm{0}   & {Z_p}^{\prime\prime\prime}\\
{X_v}^{\prime\prime\prime} & \bm{0}
\end{pmatrix}=
\begin{pmatrix}
\bm{0} &{Z_p}^{\prime\prime\prime}_{R_1} \\
{X_v}_{R_2}^{\prime \prime\prime}& \bm{0}\\
{X_v}_{R_1}^{\prime\prime\prime}& \bm{0}\\
\bm{0}   & {Z_p}^{\prime\prime\prime}_{R_2}
\end{pmatrix}
=\begin{pmatrix}
A \\ B
\end{pmatrix},
\end{equation}
where $Q$ is defined in Eq.~(\ref{eq:hadamard_conjugation}). It remains to
convert $B$ to a $2L^2\times 2L^2$ identity matrix, which is accomplished by 
appropriate column permutations. The columns of $B$, expressed as $b^{ijd}$,
are $v^{ij\prime\prime\prime}_{R_1}\oplus\bm{0}_{R_2}$ and 
$\bm{0}_{R_1}\oplus {p^{ij}_{lmd}}_{R_2}^{\prime\prime\prime}$, where 
according to Eqs.~(\ref{eq:vprimeprimeprime}) and (\ref{eq:pprimeprimeprime}) 
\begin{eqnarray}
{v^{ij}_{lmd}}_{R_1}^{\prime\prime\prime}&=&\delta_{j,m}\times
\begin{cases}
\delta_{i,l}\delta_{d,y} & i=1\\
\delta_{i-1,l}\delta_{d,x} & i\geq 2;\\
\end{cases}
\nonumber \\
{p^{ij}_{lmd}}_{R_2}^{\prime\prime\prime}&=&\delta_{j,m}\times\begin{cases}
\delta_{i+1,l}\delta_{d,y} & i<L\\
\delta_{l,L}\delta_{d,x} & i=L,
\end{cases}
\end{eqnarray}
and these are to be converted to the matrix elements of the identity
\begin{equation}
\label{eq:b^ij_lm}
b^{ijd_1}_{lmd_2} = \delta_{i,l}\delta_{j,m}\delta_{d_1,d_2}, 
\end{equation}
where $i,j,l,m\in\set{1,\ldots,L}$ and $d_1,d_2\in\set{x,y}$.

When $i\in\set{2,\ldots,L}$, the column vector of $B$ is
\begin{equation}
\begin{aligned}
v^{ij\prime\prime\prime}_{R_1}\oplus \bm{0}_{R_2} 
=&\bigoplus_{lmd\in R_1} \delta_{i-1,l}\delta_{j,m}\delta_{d,x} \oplus\bigoplus_{lmd\in R_2} 0\\
=& \bigoplus_{lmx} \delta_{i-1,l}\delta_{j,m},
\end{aligned}
\end{equation}
where the only nonzero entry is at row $(i-1,j,x)$. To map to the form in 
Eq.~(\ref{eq:b^ij_lm}), relabel this column vector:
\begin{equation}
b^{ijx}=v^{i+1,j\prime\prime\prime}_{R_1}\oplus \bm{0}_{R_2}
=\bigoplus_{lmd}\delta_{i,l}\delta_{j,m}\delta_{d,x},
\label{eq:b^ijx}
\end{equation}
where now $i\in\set{1,\ldots,L-1}$. Any column permutation on $B$ must also be 
performed on $A$. Recall from Eqs.~(\ref{eq:vprimeprimeprime}) and 
(\ref{eq:pprimeprimeprime}) that the colums of $A$, expressed as $a^{ijd}$, are 
expressed as $p^{ij\prime\prime\prime}_{R_1}\oplus \bm{0}_{R_2}$ and
$\bm{0}_{R_1}\oplus v^{ij\prime\prime\prime}_{R_2}$, with elements
\begin{eqnarray}
{p^{ij}_{lmd}}_{R_1}^{\prime\prime\prime}&=&\begin{cases}
\theta_{l,i}(\delta_{j,m}+\delta_{j+1,m})\delta_{d,x}
+\delta_{1,l}\delta_{j,m}\delta_{d,y} & i<L\\
\theta_{l,L-1}\delta_{j,m}\delta_{d,x} & i=L;
\end{cases}\nonumber \\
{v^{ij}_{lmd} }_{R_2}^{\prime\prime\prime}&=&
\begin{cases}
\theta_{2,l}\delta_{j,m}\delta_{d,y} & i=1\\
\delta_{L,l}\delta_{j,m}\delta_{d,x}
+\theta_{i,l}\left(\delta_{j,m}+\delta_{j-1,m}\right)\delta_{d,y} & i\geq 2.\\
\end{cases}\nonumber \\
\end{eqnarray}
From Eq.~(\ref{eq:QS=AB}), the column vector in $A$ with the same column index 
as $v^{ij\prime\prime\prime}_{R_1}\oplus \bm{0}_{R_2}$ is
$\bm{0}_{R_1}\oplus v^{ij\prime\prime\prime}_{R_2}$. Then
\begin{eqnarray}
a^{ijx}&=& \bm{0}_{R_1}\oplus {v^{i+1,j}_{R_2} }^{\prime\prime\prime}
\nonumber \\
&=&\bigoplus_{lmd}  \delta_{L,l}\delta_{j,m}\delta_{d,x}+
\theta_{i+1,l}\left(\delta_{j,m}+\delta_{j-1,m}\right)\delta_{d,y},
\hphantom{aa}
\label{eq:a^ijx}
\end{eqnarray}
again with $i\in\set{1,\ldots,L-1}$. Likewise, when $i=L$:
\begin{eqnarray}
b^{Ljx}&=&\bm{0}_{R_1}\oplus {p^{Lj}_{R_2}}^{\prime\prime\prime} 
= \bigoplus_{lmd} \delta_{l,L}\delta_{j,m}\delta_{d,x},
\label{eq:b^Ljx}\\
a^{Ljx}&=&{p^{Lj}_{R_1}}^{\prime\prime\prime} \oplus \bm{0}_{R_2}
=\bigoplus_{lmd} \theta_{l,L-1}\delta_{j,m}\delta_{d,x};
\label{eq:a^Ljx}
\end{eqnarray}
when $i\in\set{2,\ldots,L}$:
\begin{eqnarray}
b^{ijy}&=& \bm{0}_{R_1}\oplus{p^{i-1,j}_{R_2}}^{\prime\prime\prime} 
=\bigoplus_{lmd}  \delta_{i,l}\delta_{j,m}\delta_{d,y},
\label{eq:b^ijy}\\
a^{ijy}&=& {p^{i-1,j}_{R_1}}^{\prime\prime\prime} \oplus \bm{0}_{R_2}
\nonumber\\
&=&\bigoplus_{lmd} \theta_{l,i-1}(\delta_{j,m}+\delta_{j+1,m})\delta_{d,x}
+\delta_{1,l}\delta_{j,m}\delta_{d,y};\hphantom{aa}
\label{eq:a^ijy}
\end{eqnarray}
and when $i=1$:
\begin{eqnarray}
b^{1jy}&=& {v^{1j}_{R_1} }^{\prime\prime\prime}\oplus \bm{0}_{R_2} 
=\bigoplus_{lmd}  \delta_{1,l}\delta_{j,m}\delta_{d,y},
\label{eq:b^1jy}\\
a^{1jy}&=&\bm{0}_{R_1}\oplus{v^{1j}_{R_2} }^{\prime\prime\prime} 
=\bigoplus_{lmd}  \theta_{2,l}\delta_{j,m}\delta_{d,y}.
\label{eq:a^1jy}
\end{eqnarray}
That $B$ is an identity matrix after column permutations is clear from
Eqs.~(\ref{eq:b^ijx}), (\ref{eq:b^Ljx}), (\ref{eq:b^ijy}), and 
(\ref{eq:b^1jy}).

\begin{figure}[t]
\centering
\includegraphics[width=0.9\columnwidth]{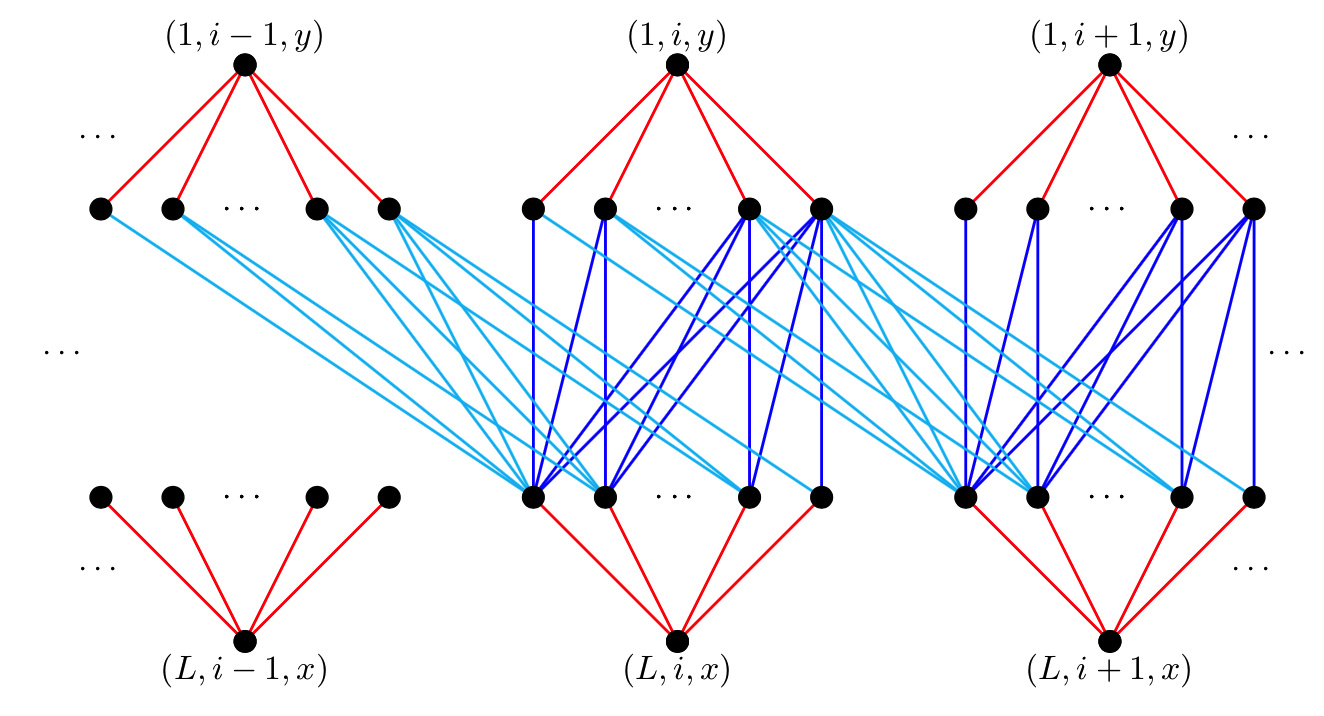}
\caption{(Color online) A portion of the toric graph. Shown are the multiple 
star graphs with 
central vertices labeled by $(1,j,y)$ and $(L,j,x)$ qubits, each connected 
to the remaining $L-1$ vertices labeled by $(i,j,y)$ ($i\in[2,L]$) and 
$(i',j,x)$ ($i'\in[1,L-1]$), respectively. Also shown are two of the half1 
graphs that connect the $(i,j,y)$ and $(i',j,x)$ vertices, and two of the half2 
graphs that connect the $(i,j,y)$ and $(i',j,x)$ vertices.}
    \label{fig:toric_graph_state_2D}
\end{figure}

Finally, the adjacency matrix $A$ for the toric code graph is obtained by 
combining Eqs.~(\ref{eq:a^ijx}), (\ref{eq:a^Ljx}), (\ref{eq:a^ijy}), and 
(\ref{eq:a^1jy}):
\begin{eqnarray}
a_{lmd_2}^{ijd_1}&=&\delta_{d_1,x}\delta_{d_2,x}\delta_{m,j}
(\delta_{l,L}\theta_{i,L-1}+\delta_{i,L}\theta_{l,L-1})\nonumber \\
&+&\delta_{d_1,y}\delta_{d_2,y}\delta_{m,j}
(\delta_{i,1}\theta_{2,l}+\delta_{l,1}\theta_{2,i})\nonumber \\
&+&\delta_{d_1,y}\delta_{d_2,x}(\delta_{m,j}
+\delta_{m-1,j})\theta_{l,i-1}\theta_{2,i}\nonumber \\
&+&\delta_{d_1,x}\delta_{d_2,y}(\delta_{m,j}
+\delta_{j-1,m})\theta_{i+1,l}\theta_{i,L-1}.
\label{eq:toric_adjacency_matrix}
\end{eqnarray}
Note that the matrix elements in the expression above are symmetric under 
$i\leftrightarrow l,j\leftrightarrow m$ and $d_1\leftrightarrow d_2$ ; the apparent lack of symmetry
in the last two lines is resolved by noting that
\begin{align}
\theta_{i+1,l}\theta_{i,L-1}=& \theta_{i,l-1}\theta_{2,l}\theta_{i,L-1}
=\theta_{i,l-1}\theta_{2,l}.
\end{align}
Eq.~(\ref{eq:toric_adjacency_matrix}) is the first of two key results of the 
present work. The graph represented by the adjacency matrix will be referred to
as the toric graph, illustrated in Fig~\ref{fig:toric_graph_state_2D}, and the
corresponding graph state is called the toric graph state.

The result~(\ref{eq:toric_adjacency_matrix}) was checked in two ways. First, 
the graph state for small systems was generated explicitly and compared with
the toric code state on the same number of qubits. Second, the reduced 
density matrices and entanglement entropies for various bipartitions of the 
two systems were compared and found to agree in all cases.

\subsection{Decomposition of the toric graph}

The elements of the graph adjacency matrix, 
Eq.~(\ref{eq:toric_adjacency_matrix}), and the associated graph shown in 
Fig~\ref{fig:toric_graph_state_2D}, appear too complicated to gain any insights 
about why this particular structure corresponds to a topological quantum state. 
However, it turns out this graph can be decomposed into three subgraphs, all of 
which have a rather simple structure. In particular, the adjacency matrix given 
can be decomposed into three terms:
\begin{equation}
\label{eq:decomposition_toric_adjacency_matrix}
A = A^{\text{mstar}} + A^{\text{mhalf1}} +A^{\text{mhalf2}},
\end{equation}
where the entries of those three matrices are 
\begin{eqnarray}
A^{\text{mstar}}_{i_1j_1d_1,i_2j_2d_2} 
&=&\delta_{d_1,y}\delta_{d_2,y}\delta_{j_1,j_2}( \delta_{i_2,1}\theta_{2,i_1}
+\delta_{i_1,1}\theta_{2,i_2} )\nonumber \\
&+& \delta_{d_1,x}\delta_{d_2,x}\delta_{j_1,j_2}(\delta_{i_1,L}
\theta_{i_2,L-1}+\delta_{i_2,L}\theta_{i_1,L-1} );\nonumber \\
A^{\text{mhalf1}}_{i_1j_1d_1,i_2j_2d_2} 
&=& \delta_{d_1,x}\delta_{d_2,y}\delta_{j_1,j_2}\theta_{i_1,i_2-1}
\theta_{2,i_2}\nonumber \\
&+&\delta_{d_1,y}\delta_{d_2,x}\delta_{j_1,j_2}\theta_{i_2,i_1-1}
\theta_{2,i_1};\nonumber \\
A^{\text{mhalf2}}_{i_1j_1d_1,i_2j_2d_2} 
&=& \delta_{d_1,x}\delta_{d_2,y}\delta_{j_1-1,j_2}\theta_{i_1,i_2-1}
\theta_{2,i_2}\nonumber \\
&+&\delta_{d_1,y}\delta_{d_2,x}\delta_{j_2-1,j_1}\theta_{i_2,i_1-1}
\theta_{2,i_1}.
\label{eq:A^mhalf2}
\end{eqnarray}
These three matrices corresponding to three subgraphs $G_{\text{mstar}}, 
G_\text{mhalf1}$, and $G_\text{mhalf2}$, respectively.   

\begin{figure}[t]
\centering
\includegraphics[width=0.48\textwidth]{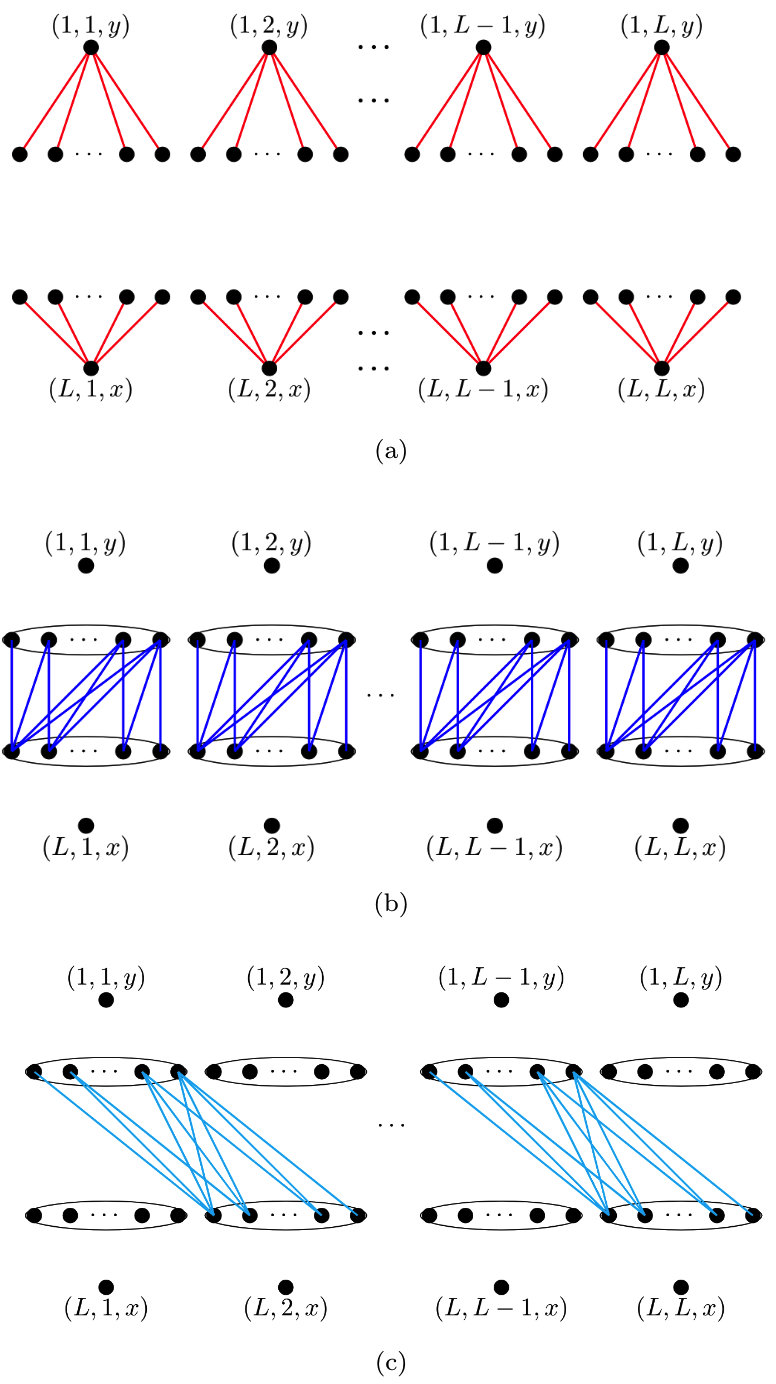}
\caption{(Color online) The three subgraphs of the toric graph: (a) $G_{\text{mstar}}$, (b) $G_{\text{mhalf1}}$, (c)$G_{\text{mhalf2}}$.}
\label{fig:toric_subgraph}
\end{figure}

First consider the subgraph $G_{\text{mstar}}$ and the subset of vertices
\begin{equation}
Q^1_{jd}:=\set{1,\ldots,L}_d\times\set{j}_d,
\end{equation}
following the notation of Eq.~(\ref{eq:row_index_xz}). Because of the 
restriction $\delta_{j_1,j_2}$ in the definition of $A^{\rm mstar}$ above,
there is no edge in $G_{\text{mstar}}$ connecting qubits in different subsets 
$Q^1_{jd}$. There are $2L$ disconnected components in total, and the adjacency 
matrices of components $Q^1_{jx}$ and $Q^1_{jy}$ have elements
\begin{eqnarray}
(A_{jx}^\text{star})_{i_1,i_2}=A^{\text{star}}_{i_1jx,i_2jx}
&=&\delta_{i_1,L}\theta_{i_2,L-1}+  \delta_{i_2,L}\theta_{i_1,L-1};\nonumber \\
(A_{jy}^\text{star})_{i_1,i_2}=A^{\text{star}}_{i_1jy,i_2jy}
&=&\delta_{i_2,1}\theta_{2,i_1}+\delta_{i_1,1}\theta_{2,i_2}
\label{eq:star_adjacency_jd_component}
\end{eqnarray}
respectively.
From Eq.~(\ref{eq:A^star_ij}), the induced subgraphs on $Q^1_{jx}$ and 
$Q^1_{jy}$ are $L$-vertex star graphs with $(L,j,x)$ and $(1,j,y)$ being the 
central vertices, respectively, as shown in Fig~\ref{fig:toric_subgraph}(a). Thus, 
`mstar' is an abbreviation for `multiple star graphs.'

The graph $G_{\text{mhalf1}}$ is similarly made up of $L$ disconnected
components, indexed by $j$. The adjacency matrix elements are
\begin{eqnarray}
\label{eq:adjacency_matrix_half_j}
({A_{j}^\text{mhalf1}})_{i_1d_1,i_2d_2} 
&=& A^{\text{mhalf1}}_{i_1jd_1,i_2jd_2} \nonumber\\
&=& \delta_{d_1,x}\delta_{d_2,y}\theta_{i_1,i_2-1}\theta_{2,i_2}\nonumber \\
&+& \delta_{d_1,y}\delta_{d_2,x}\theta_{i_2,i_1-1}\theta_{2,i_1}.
\end{eqnarray}
When $i_1=1$, $\theta_{2,i_1}$ is always zero, so vertex $(1,j,y)$ is isolated;
likewise $\theta_{i_1,i_2-1}\theta_{2,i_2}$ is always zero when $i_1=L$ and
$(L,j,x)$ is also isolated.
Based on Eqs.~(\ref{eq:adjacency_matrix_half_j}) and (\ref{eq:half_A_xiyj}), 
$G_{\text{mhalf1}}$ corresponds to multiple copies of a $2(L-1)$-vertex half 
graph, as shown in Fig.~\ref{fig:toric_subgraph}().

$G_{\text{mhalf2}}$ is almost the same as $G_{\text{mhalf1}}$: composed of $L$ 
disconnected components and each component is a $2(L-1)$-vertex half graph. The 
only difference is that the vertices in each of the components are different:
\begin{equation}
Q^2_j:= \set{1,\ldots,L}_d\times\left(\set{j}_y\cup\set{j+1}_x\right).    
\end{equation}
The adjacency matrix of induced subgraph on each component $Q^2_j$ is the same 
as in Eq.~(\ref{eq:adjacency_matrix_half_j}), and the graph is shown in 
Fig~\ref{fig:toric_subgraph}(c).

\subsection{Observations on the toric graph structure}
\label{sec:obser_toric_graph}


As discussed in Sec.~\ref{sec:stabilizer_and_graph_state}, the stabilizer state 
represented by the star graph is LC-equivalent to the GHZ state, so the graph 
state represented by $G_{\text{mstar}}$ is LC-equivalent to multiple copies of
the GHZ state. In fact, such a multiple-copy GHZ state is already 
`topologically ordered,' in the sense that it is a code state in a quantum 
error correction code with macroscopic distance $d\sim \sqrt{N}$. Consider two 
$m$-qubit GHZ states
\begin{align}
\Ket{\phi_m^+} = \frac{\Ket{0^m}+\Ket{1^m}}{\sqrt{2}},\quad
\Ket{\phi_m^-} = \frac{\Ket{0^m}-\Ket{1^m}}{\sqrt{2}}.
\end{align}
and their $m$-copy states on $m^2$ qubits
\begin{equation}
\label{eq:m_copy_GHZ}
\Ket{\varphi^+} =  \Ket{\phi_m^+}^{\otimes m},
\Ket{\varphi^-} =  \Ket{\phi_m^-}^{\otimes m}.
\end{equation}
Then, $\operatorname{span}_{\mathbb{C}}\Set{\Ket{\varphi^+},
\Ket{\varphi^-}}$ is a quantum error correction code with distance $d=m$. The 
proof is given in Appendix~\ref{sec:proof}. The reader might recognize that 
when $m=3$, the code $\operatorname{span}_{\mathbb{C}}\Set{\Ket{\varphi^+},
\Ket{\varphi^-}}$ is nothing but Shor's celebrated nine-qubit (repetition) 
code~\cite{shor1995,nielsen2000}. The decomposition of the toric graph thus 
reveals an intriguing and apparently novel connection between the toric code 
and the repetition code. 

The close connection between multiple GHZ states and the toric code is perhaps
surprising. On the one hand, the GHZ states represent the long-range 
entanglement exhibited by topological states, spanning the length of the 
system. On the other hand, GHZ states are the most fragile many-qubit entangled
states; a single measurement of any of the constituent qubits deletes all of 
the edges within the star graph. 
Because of the translational invariance, however, the resulting toric graph maintains the 
same connectivity on the remaining qubits. Thus, the toric graph is effectively 
invariant under single-qubit measurements, demonstrating the robustness of the 
underlying topology.

Moreover, the multi-copy GHZ state is also a simple example of the 
distance balancing technique proposed in Ref.~\cite{hastings2016weight}.
Consider the code $\operatorname{span}_{\mathbb{C}}\{\Ket{\phi_m^+},
\Ket{\phi_m^-}\}$, in which $d_X=m$ and $d_Z=1$, where $d_X$ and $d_Z$ are the
distances with respect to X and Z errors, respectively. The distance balance 
method takes $d_X/d_Z$ copies of such a code and outputs a new code with 
distance $\Tilde{d}_X=\tilde{d}_Z=d_X$, at the cost of increasing the number of 
physical qubits. By adding another layer of structure corresponding to the half 
graphs, the degree of many vertices is not still bounded, yet the weight of the
original stabilizer generators in the original (toric) code remains constant.
The star graphs and half graphs thus play different key roles: the multi-star 
graphs contribute the large code distance while the half graphs ensure local
stabilizer generators.

There is also a close resemblance between the code 
$\operatorname{span}_{\mathbb{C}}\Set{\Ket{\varphi^+},\Ket{\varphi^-}}$ and the 
recently proposed repetition cat code~\cite{Guillaud2019Repetition} in the 
continuous variable setting, where two approximately orthogonal coherent states 
are used as the qubit registers $\Ket{0}_c=\Ket{+\alpha}$ and 
$\Ket{1}_c=\Ket{-\alpha}$. Similar to 
$\operatorname{span}_{\mathbb{C}}\{\Ket{\phi_m^+},\Ket{\phi_m^-}\}$, the bit 
flip error is exponentially suppressed $(d_X=m)$ while the phase error is 
likely to occur $(d_Z=1)$. One therefore defines the repetition cat qubit state
as $\Ket{\pm}_L =\Ket{\pm}_c^{\otimes r}$ in order to  correct the phase error, 
which corresponds to 
$\operatorname{span}_{\mathbb{C}}\Set{\Ket{\varphi^+},\Ket{\varphi^-}}$ if 
$r=m$.

\section{Toric graph state generation}
\label{sec:generation_toric_graph_state}

This section focuses on how to generate toric code states, and encode arbitrary
quantum states in the toric QECC, within the quantum circuit model. We show
that this can be accomplished in log depth in the absence of ancillae, and in 
constant depth including ancillae. The key step is to construct log-depth 
quantum circuits that generate the two toric code subgraphs: the star and half 
graphs.

The goal is to prepare the quantum state
\begin{equation}
\label{eq:f_quantum_state}
\Ket{f} = \frac{1}{\sqrt{2^N}} \sum_{q\in\set{0,1}^N} (-1)^{f(q)}\Ket{q}     
\end{equation}
where 
the Boolean function $f:\set{0,1}^N\to\set{0,1}$ is associated with the 
binary quadratic form for a graph, Eq.~(\ref{eq:graph_state_explicit_form}).
Given an operator $U_f$, which implements 
\begin{equation}
\label{eq:unitary_phase}
U_f\Ket{q}=(-1)^{f(q)}\Ket{q},\quad \forall q\in\set{0,1}^N, 
\end{equation}
one has $\Ket{f}= U_{f}H^{\otimes N}\Ket{0^N}$. Moreover, if $f$ can be 
decomposed into the sum (module 2) of other Boolean functions, i.e.\ 
$f(q) = f_1(q)+\cdots+f_k(q)$, then $\Ket{f}$ can be generated by applying the 
commuting $U_{f_i}$ operators in sequence:
\begin{equation}
\label{eq:composition_qunatum_gate}
\Ket{G}= U_f^{(k)}\cdots U_f^{(1)}H^{\otimes N}\Ket{0^N}.
\end{equation}

The target is the toric graph state
\begin{equation}
 \Ket{G_\text{toric}} = \frac{1}{\sqrt{2^{2L^2}}}\sum_{q\in\set{0,1}^{2L^2}}(-1)^{f_\text{toric}(q)}\Ket{q},
\end{equation}
where $f_\text{toric}(q)
=\sum_{n_1<n_2 } q_{n_1}q_{n_2}(A_{\text{toric}})_{n_1,n_2}$ is the quadratic 
Boolean form related to toric graph and
\begin{equation}
(A_{\text{toric}})_{n_1,n_2}=(A_{\text{toric}})_{i_1j_1d_1,i_2j_2d_2}
\end{equation}
are the matrix elements of the graph adjacency matrix, 
Eq.~(\ref{eq:toric_adjacency_matrix}). The variables 
$n_1,n_2\in\set{1,\ldots,2L^2}$ are mapped to the qubit lattice coordinates 
$ijd$ via
\begin{equation}
(i,j,d)\leftrightarrow i + (j-1)L + \delta_{d,y}L^2.
\end{equation}

Based on the decomposition of the toric graph and the relation between 
quadratic Boolean forms and the graph adjacency matrices, $f_\text{toric}$ can 
be decomposed into 
\begin{eqnarray}
f_\text{toric}&=&f_\text{mstar}+f_\text{mhalf1}+f_\text{mhalf2}\nonumber \\
&=&\sum_{jd} f^\text{star}_{jd}+\sum_{j}f^\text{half1}_{j}
+\sum_{j}f^\text{half2}_{j},
\end{eqnarray}
where the adjacency matrices for $f^\text{star}_{jd}$, $f^\text{half1}_{j}$, 
and $f^\text{half2}_{j}$ are defined in 
Eqs.~(\ref{eq:star_adjacency_jd_component}) and 
(\ref{eq:adjacency_matrix_half_j}), respectively. Thus, $\Ket{G_\text{toric}}$ 
can be generated by the following circuits
\begin{equation}
\ket{G_\text{toric}} = 
U^{\rm mhalf}U^{\rm mstar} H^{\otimes 2L^2}\Ket{0}^{\otimes 2L^2},
\label{eq:qcircuit_toric_graph_state}
\end{equation}
where
\begin{align}
\label{eq: U-mhalf}
U^{\rm mhalf} =& \prod_j U_{f,j}^{\rm half2} \prod_j U_{f,j}^{\text{half1}}, \\
U^{\rm mstar} =& \prod_{jd}U_{f,jd}^{\text{star}}.
\end{align}
Different $U_{f,jd}^{\text{star}}$ operators have the same circuit depth, as 
they all compute the same quadratic Boolean function associated with an 
$L$-vertex star graph; moreover, each acts on different subsets of qubits, so 
all can be performed in parallel. The situation is similar for 
$U_{f,j}^{\text{half1}}$ and $U_{f,j}^{\text{half2}}$. To summarize: the depth 
of the toric code quantum circuit in Eq.~(\ref{eq:qcircuit_toric_graph_state}) 
corresponds to the sum of the circuit depths for $U_{f,1x}^{\rm star}$,
$U_{f,1}^{\rm half1}$ and $U_{f,1}^{\rm half2}$.

\subsection{Generation of the star graph state}
\label{sec:generate_star_graph_state}

The $f^\text{star}_{jd}$ corresponds to the quadratic form for each $L$-vertex 
star graph, which is generically expressed as 
$f(q)=\left(q_1+\cdots +q_{c-1}+q_{c+1}+\cdots q_L\right)q_c$ with $c$ labeling
the high-degree central vertex, the value of which is unimportant. The term in 
parentheses corresponds to the parity $\operatorname{Parity}(q_1,\ldots,q_n)$
of an $n$-length string $q_1\oplus\cdots \oplus q_n$. The log-depth quantum 
circuit is inspired by the classical parity algorithm in a parallel setting. 
First, divide all $n$ elements into $n/2$ disjoint pairs and calculate the 
parity of each subset in parallel; then continue subdividing until only one 
pair remains. It requires $\log_2n$ iterations to obtain the parity of the 
bitstring (this of course ignores the $n$ bits of classical communication 
required). In the quantum setting, the parity doesn't need to be calculated; 
only the two-qubit gates need to be implemented that generate the appropriate 
contribution to the Boolean function. The operations at each iteration commute, 
and therefore they can be truly implemented in parallel.

Without loss the generality, suppose the central vertex is the last one, and 
the unitary operation implementing Eq.~(\ref{eq:unitary_phase}) with 
$f(q)=\left(q_1+\cdots+\cdots q_{L-1}\right)q_L$ is
\begin{equation}
\label{eq:U_f_s}
U_f^{\rm star}=P^{-1}\operatorname{CZ}(q_{L-1},q_L)P,    
\end{equation}
in which $P$ is a quantum operation that generates the parity linear form:
\begin{equation}
\label{eq:gate_P}
P\Ket{q}=\Ket{q^\prime},
\end{equation}
where $q^\prime_{L-1} = q_1+\cdots+q_{L-1}$. The operator $P$ can be 
implemented using a series of $\operatorname{CX}=\operatorname{CNOT}$ gates, 
which have the action
\begin{equation}
\operatorname{CX}(1,2)\Ket{q_1,q_2}=\Ket{q_1,q_1+q_2}.    
\end{equation}
One first divides the first $L-1$ qubits into $(L-1)/2$ disjoint pairs and 
calculates the parity of each pair in parallel:
\begin{equation}
\prod_{i=1}^{\lfloor (L-1)/2 \rfloor}\operatorname{CX}(2i-1,2i)\Ket{q}
=\Ket{\tilde{q}},
\end{equation}
where $\Ket{\tilde{q}}$ denotes the state in quantum register after first 
iteration and $\tilde{q}_{2i} = q_{2i-1}+q_{2i}$. If $L-1$ is odd, then the 
$L-1$-th qubit does not need to be explicitly paired. One then divides the 
$\lceil \frac{L-1}{2}\rceil$ quantum registers with the parity result to 
$(L-1)/4$ pairs, and repeats the procedure until all of the clauses have been
paired. After $\log(L-1)$ iterations, one obtains the result in the $L-1$-th 
register as $q^\prime_{L-1} = q_1+\cdots+q_{L-1}$. Next, the
$\operatorname{CZ}$ gate implements the required phase:
\begin{equation}
\begin{aligned}
\operatorname{CZ}(L-1,L)\Ket{q^\prime} 
=& (-1)^{q^\prime_{L-1}\cdot q^\prime_{L}}\Ket{q^\prime} \\
=& (-1)^{(q_1+\cdots+q_{L-1})q_L}\Ket{q^\prime}.
\end{aligned}
\end{equation} 
As $U_f^{\rm star}$ should yield $(-1)^{f(q)}\Ket{q}$ as the output state, one 
must implement the inverse of $P$ to change $\Ket{q^\prime}$ to $\Ket{q}$:
\begin{equation}
P^{-1}(-1)^{f(q)}\Ket{q^\prime} = (-1)^{f(q)}\Ket{q}.    
\end{equation}

\begin{figure}[t]
\centering
\includegraphics[width=0.8\columnwidth]{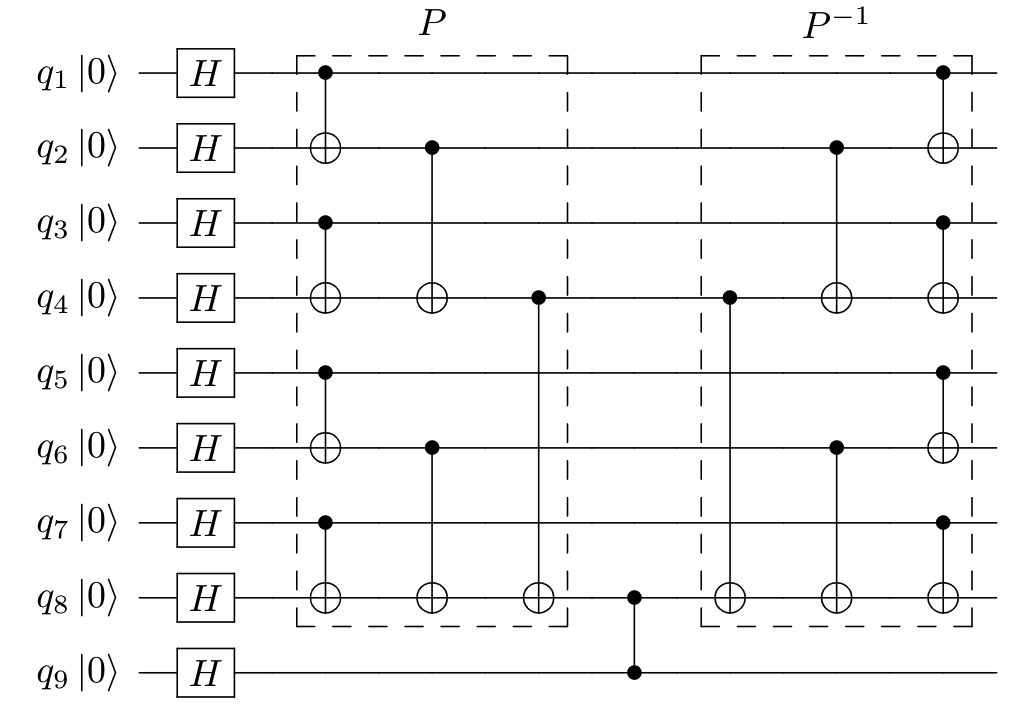}
\caption{Quantum circuit that generates a nine-qubit star graph state}
\label{fig:qcircuit_star_graph}
\end{figure}

The nine-qubit example for $U_f^{\rm star}$ is shown in Fig.~\ref{fig:qcircuit_star_graph}. 
The construction of $P$ for arbitrary 
number of qubits, Algorithm~\ref{alg:generate_star_graph_state}, and the proof 
of its log depth are given in 
Appendix~\ref{sec:appen_algirothm_state_graph_state}. Thus, the depth of 
$U_{f,1x}^{\rm star}$ is $O(\log L)$. Note that a log-depth circuit for the 
realization of GHZ states has been obtained recently by other 
means~\cite{Cruz2019}.

\subsection{Generation of the half graph state}
\label{subsec:half graph}

The Boolean quadratic forms $f^\text{half1}_{j}$ and $f^\text{half2}_{j}$ are 
associated with the $2(L-1)$-vertex half graph, so the current task is to 
construct a quantum circuit that computes $f_h$. For a $2n$-vertex half graph, 
$f_h(q,p)=\sum_{i\le j}q_ip_j$, where $i,j\in \set{1,\ldots,n}$.
Moreover, $f_h(q,p)$ can be decomposed in the following way:
\begin{equation}
f_h(q,p) = \sum_{k=0}^{\lceil\log n\rceil} f_h^{(k)}(q,p),  
\label{eq:decomposition_half_boolean_function}
\end{equation}
where
\begin{equation}
\label{eq:f_h^k}
\begin{aligned}
&f_h^{(k)}(q,p)\\
&=\sum_{i=0}^{n/2^k-1}\left(\sum_{j=2^ki+1}^{2^k(i+\frac{1}{2})} q_j \right)
\left(\sum_{j=2^k(i+\frac{3}{2})}^{2^k(i+1)} p_j \right) 
\end{aligned}
\end{equation}
assuming that $n$ is a power of two. Otherwise, one need only replace the sum 
upper bound by $n/2^k \to \lceil n/2^k\rceil$, 
$2^k(i+1/2)\to \min\{2^k(i+1/2),n\}$, and $2^k(i+1)\to \min\{2^k(i+1),n\}$.
For simplicity of analysis one can assume that $n$ is a power of two, but the 
results hold for arbitrary integer values. A few decomposed Boolean functions 
are listed as follows:
\begin{eqnarray}
f_h^{(0)}(q,p)&=& \sum_{i=0}^{n-1}q_{i+1}p_{i+1};\nonumber \\
f_h^{(1)}(q,p)&=& \sum_{i=0}^{n/2-1}q_{2i+1}p_{2i+2};
\nonumber \\
f_h^{(2)}(q,p)&=& \sum_{i=0}^{n/4-1}(q_{4i+1}+q_{4i+2})(p_{4i+3}+p_{4i+4});
\nonumber \\
f_h^{(\log_2n)}(q,p)&=&\left(\sum_{j=1}^{n/2}q_j\right)
\left(\sum_{j=n/2+1}^np_j\right),  
\end{eqnarray}

\begin{figure}[t]
    \centering
    \includegraphics[width=0.8\columnwidth]{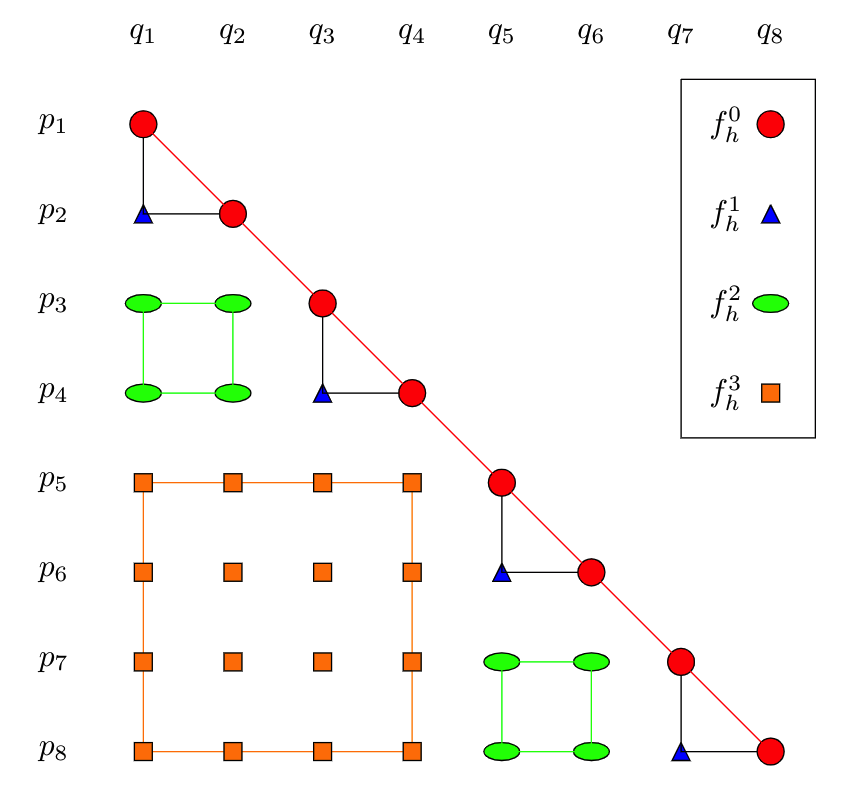}
    \caption{(Color online) Decomposition of $f_h(q,p)$ when $n=2^3$.}
    \label{fig:half_boolean_decomp}
\end{figure}

The decomposition for $f_h(q,p)$ when $n=2^3$ is shown in 
Fig~\ref{fig:half_boolean_decomp}, from which one may obtain an intuition of
why the decomposition Eq.~(\ref{eq:decomposition_half_boolean_function}) holds 
in general. In Fig~\ref{fig:half_boolean_decomp}, the columns correspond to 
variable $q_i$ and the rows correspond to variable $p_i$. Each element 
represents a term $q_ip_j$ appearing in $f_h(q,p)$. From the condition $i\le j$ 
in the sum of $f_h(q,p)$, one obtains the representation as a triangle. The 
decomposition Eq.~(\ref{eq:decomposition_half_boolean_function}) corresponds to 
separating the triangle into a square and two triangles of half size 
iteratively. For example, the $2^3\times 2^3$ triangle in 
Fig~\ref{fig:half_boolean_decomp} is decomposed into the left bottom 
$2^2\times 2^2$ square (corresponding to $f_h^{(3)}$) and two $2^2\times 2^2$ 
triangles above and on the right of it. Decomposing these triangles in turn 
yields $2\times 2$ square (corresponding to the $f_h^{(2)}$), with the 
$f_h^{(1)}$ and $f_h^{(0)}$ terms remaining.

Each square corresponds to a term in the Boolean function of the form 
$(\sum_i q_i)(\sum_j p_j)$, where  $(\sum_i q_i)$ again corresponds to the 
parity operation which can be implemented in log depth. The 
sum in Eq.~(\ref{eq:f_h^k}) consists of $n/2^k$ squares of
side length $2^{k-1}$. As different squares associated to the same $k$ share 
no common variables, their parity operations can be implemented in parallel.
Combining Eqs.~(\ref{eq:composition_qunatum_gate}) and
(\ref{eq:decomposition_half_boolean_function}), the half graph operators can 
therefore be implemented with a $O(\log^2n)$-depth quantum circuit.

This circuit depth can be further reduced by a more careful construction. Note
that one of the sums in the penultimate term of 
Eq.~(\ref{eq:decomposition_half_boolean_function})
\begin{eqnarray}
f_h^{(\log_2n-1)}(q,p)&=&\left(\sum_{j=1}^{n/4}q_j\right)
\left(\sum_{j=n/4+1}^{n/2}p_j\right)\nonumber \\
&+&\left(\sum_{j=n/2+1}^{3n/4}q_j\right)
\left(\sum_{j=3n/4+1}^{n}p_j\right),\hphantom{aa}
\end{eqnarray}
already includes half of the terms required by the final term 
$f_h^{(\log_2n)}(q,p)$. Thus, if one were to also include the calculation of 
$\sum_{j=1+n/4}^{n/2}q_j$ and $\sum_{j=n/2+1}^{3n/4}p_j$ at level $\log_2n-1$ (which
shares no variables with other terms at this level), the level-$\log_2n$ 
calculation would require only $O(1)$ operations (one multiplication). Applying 
this idea recursively, we can compute $f_h$ using a $O(\log n)$-depth quantum 
circuit, so the depth of $f_\text{half1}$ and $f_\text{half2}$ are both 
$O(\log L)$. 

Let's consider each term more carefully. The $f_h^{(0)}$ term is the sum of 
multiplications, all of which share no common variables, so that 
$U_{f_h^{(0)}}$ can be implemented as a depth-one quantum circuit:
\begin{equation}
U_{f_h^{(0)}}\Ket{q,p} = \prod_{i=1}^{n} \operatorname{CZ}(q_i,p_i)\Ket{q,p}   
=(-1)^{f_h^{(0)}(q,p)}\Ket{q,p}.
\end{equation}
Similarly,
\begin{equation}
\begin{aligned}
U_{f_h^{(1)}}\Ket{q,p} 
=& \prod_{i=0}^{n/2-1} \operatorname{CZ}(q_{2i+1},p_{2i+2})\Ket{q,p}   \\
=&(-1)^{f_h^{(1)}(q,p)}\Ket{q,p},
\end{aligned}
\end{equation}
is also obtained with a depth-one quantum circuit. To construct 
$U_{f_h^{(2)}}$, one requires $\operatorname{CX}$ gates because the sum term 
involves the parity of two bits:
\begin{eqnarray}
\Ket{q^{(2)},p^{(2)}}&=&\prod_{i}^{n/4-1}\operatorname{CX}(q_{4i+1},q_{4i+2})
\operatorname{CX}(q_{4i+3},q_{4i+4})\nonumber \\
&\times&\operatorname{CX}(p_{4i+1},p_{4i+2})
\operatorname{CX}(p_{4i+3},p_{4i+4})\Ket{p,q},\hphantom{aaa}
\label{eq:f_h^2_parity}
\end{eqnarray}
where 
\begin{eqnarray}
q_{4i+2}^{(2)} &=&  q_{4i+1}+q_{4i+2}, \quad
q_{4i+4}^{(2)} =  q_{4i+3}+q_{4i+4},\nonumber \\
p_{4i+2}^{(2)} &=&  p_{4i+1}+p_{4i+2},\quad
p_{4i+4}^{(2)} =  p_{4i+3}+p_{4i+4}.
\end{eqnarray}
The $\operatorname{CZ}$ operation then yields the desired phase:
\begin{eqnarray}
&&\prod_{i=0}^{n/4-1}\operatorname{CZ}(q_{4i+2},q_{4i+4})\Ket{q^{(2)},p^{(2)}} 
\nonumber \\
&=&(-1)^{\sum_{i=0}^{n/4-1} q^{(2)}_{4i+2}\cdot p^{(2)}_{4i+4} }
\Ket{q^{(2)},p^{(2)}}\nonumber \\
&=&(-1)^{f_h^{(2)}(p,q)}\Ket{q^{(2)},p^{(2)}}.
\label{eq:f_h^2_phase}
\end{eqnarray}
As the parity operator needs to also be used in the next iteration, the inverse 
of Eq.~(\ref{eq:f_h^2_parity}) is not applied immediately, but rather only
after all the phases $(-1)^{f_h^{(k)}(q,p)}$ have been added. 
Eqs.~(\ref{eq:f_h^2_parity}) and (\ref{eq:f_h^2_phase}) correspond to a 
depth-two quantum circuit which implements a phase $(-1)^{f_h^{(2)}(q,p)}$, and 
the same is true for all $k>2$. After the $\log(n)$ iteration, one obtains 
$(-1)^{f_h(q,p)}\Ket{q^\prime,p^\prime}$ by means of a quantum circuit of total 
depth $2\log(n)$. One then applies the inverse of all the $\operatorname{CX}$ 
gates so that the state in the quantum register is changed to 
$(-1)^{f_h(q,p)}\Ket{q,p}$, so $U_{f_h}$ is implemented using a 
$3\log(n)$-depth quantum circuit. 

\begin{figure}[t]
\centering
\includegraphics[width=0.8\columnwidth]{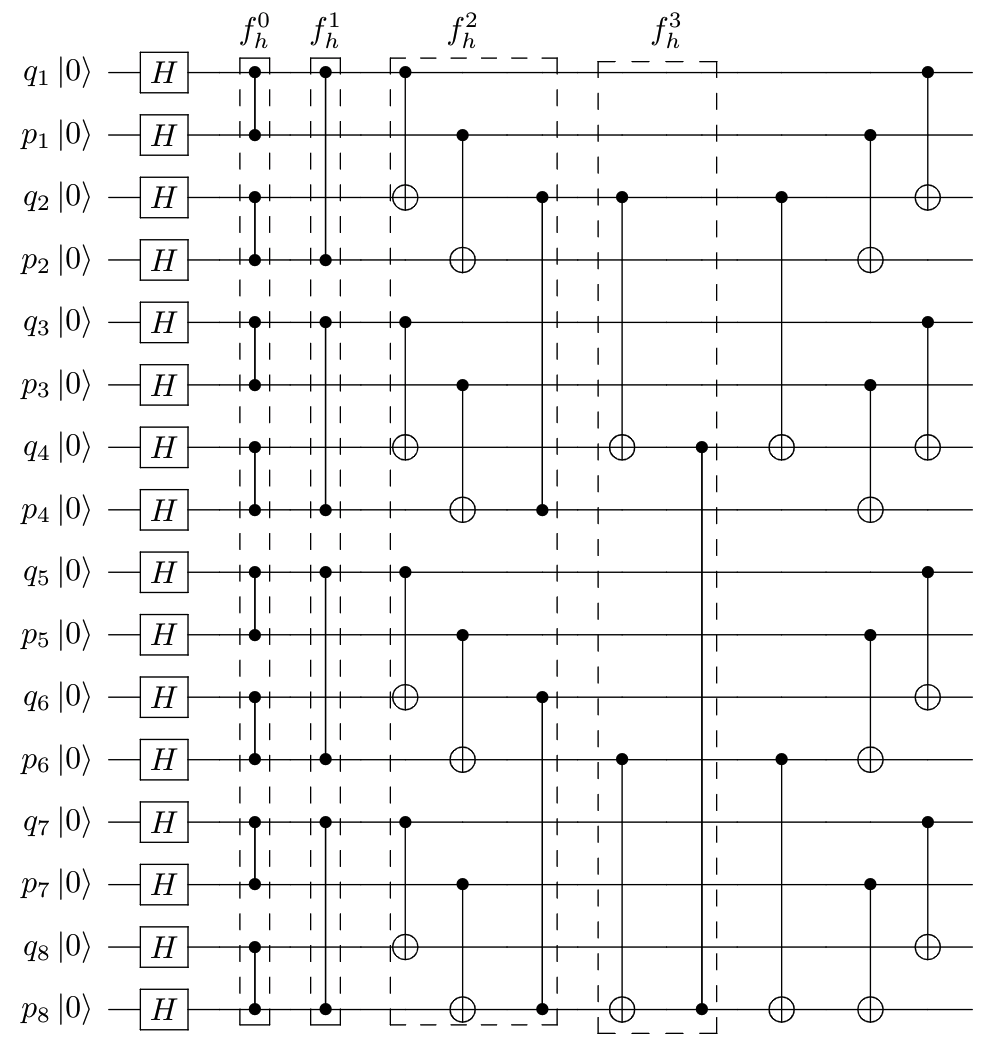}
\caption{Quantum circuit that generates a 16-qubit half graph state}
\label{fig:qcircuit_half_graph_state}
\end{figure}

The explicit circuit for the 16-qubit example is shown in 
Fig.~\ref{fig:qcircuit_half_graph_state}. 
The generic algorithm for 
arbitrary numbers of qubits, Algorithm~\ref{alg:generat_half_state}, and the 
proof of its log depth, are included in 
Appendix~\ref{sec:appendix_qcircuit_half_graph_state}. Thus, one can generate 
$\Ket{G_\text{toric}}$ via Eq.~(\ref{eq:qcircuit_toric_graph_state}) in
log depth.

\subsection{Encoding an arbitrary unknown state}

For fault-tolerant quantum computation within the toric QECC, it suffices to 
prepare only one of the degenerate states of the code, followed by 
fault-tolerant logical operations to transform the initial state to the desired 
state. However, the procedure is necessarily different if one is provided with 
an unknown quantum and asked to encode this withing the toric QECC.  In 
Ref.~\cite{Aguado2008Entanglement}, it was shown that such an encoding can be 
effected in log-depth by modifying surface code stabilizer elements.
In this section, we show that encoding arbitrary states into the
toric code can be performed in log depth using the graph state insights above.

Let $Z_\alpha:=\otimes_{i=1}^{L}Z_{(L,i,x)}$ denote the string of $Z$ operators
acting on the central vertices of star graphs on $x$ edges of the lattice, and 
$Z_\beta=\otimes_{i=1}^{L} Z_{(1,i,y)}$ defined analogously but for $y$ 
edges. The four logical states after Hadamard conjugation are 
$\Ket{G_{\rm toric}}$, $Z_\alpha\Ket{G_{\rm toric}}$, 
$Z_\beta\Ket{G_{\rm toric}}$, and $Z_\alpha Z_\beta\Ket{G_{\rm toric}}$,
where $\Ket{G_{\rm toric}}$ is defined in 
Eq.~(\ref{eq:qcircuit_toric_graph_state}). Given an unknown two-qubit state  
\begin{equation}
\label{eq:unknown_2qubit_state}
\Ket{\psi}=c_1\Ket{00}+c_2\Ket{01}+c_3\Ket{10}+c_4\Ket{11},    
\end{equation}
the aim is to prepare the encoded logical state
\begin{align}
\label{eq:target_encoding_state}
\Ket{\psi_{\rm logical}}=\left(c_1+c_2 Z_\beta+c_3Z_\alpha+c_4 Z_\alpha Z_\beta 
\right)\Ket{G_{\rm toric}}.
\end{align}

Start with 
\begin{equation}
H^{\otimes L}\Ket{0}^{\otimes(L-1)}\Ket{1} = \frac{1}{\sqrt{2^L}}\sum_{q\in\set{0,1}^L} (-1)^{q_L}\Ket{q}.
\end{equation}
After applying $U^{\rm star}$, Eq.~(\ref{eq:U_f_s}), one obtains
\begin{eqnarray}
&&\frac{1}{\sqrt{2^L}}\sum_{q\in\set{0,1}^L} (-1)^{q_L+(q_1+\cdots+q_{L-1})q_L}
\Ket{q}\nonumber \\
&=&\frac{1}{\sqrt{2^L}}\sum_{q\in\set{0,1}^{L-1}} \left(\Ket{q,0}
-(-1)^{q_1+\cdots+q_{L-1}}\Ket{q,1} \right),\hphantom{aaaa}
\end{eqnarray}
so that
\begin{equation}
\label{eq:HU*H00}
\left(H^{\otimes L-1}\otimes I\right) U^{\rm star} H^{\otimes L}
\Ket{0}^{\otimes(L-1)}\Ket{1} = \frac{\Ket{0}^{\otimes L}-\Ket{1}^{\otimes L}}
{\sqrt{2}}.
\end{equation}
Likewise,
\begin{equation}
\label{eq:HU*H01}
\left(H^{\otimes L-1}\otimes I\right) U^{\rm star} H^{\otimes L}
\Ket{0}^{\otimes(L-1)}\Ket{0} = \frac{\Ket{0}^{\otimes L}+\Ket{1}^{\otimes L}}
{\sqrt{2}}.
\end{equation}
Denote the operator $\left(H^{\otimes L-1}\otimes I\right) U^{\rm star}
\left(H^{\otimes {L-1}}\otimes I\right)$ as $U^{\rm GHZ}$. Combining
Eq.~(\ref{eq:HU*H00}) and Eq.~(\ref{eq:HU*H01}), one readily obtains
\begin{eqnarray}
U^{\rm GHZ} \Ket{0}^{\otimes L-1}\Ket{0}&=&\Ket{0}^{\otimes L};\nonumber \\
U^{\rm GHZ} \Ket{0}^{\otimes L-1}\Ket{1}&=&\Ket{1}^{\otimes L}.
\end{eqnarray}

Introducing $2(L-1)$ ancillary qubits, all initialized in state $\Ket{0}$, the
state (\ref{eq:unknown_2qubit_state}) can be written as
\begin{eqnarray}
\Ket{\psi}&=&c_1\Ket{0}^{\otimes(L-1)}\Ket{0}\Ket{0}^{\otimes(L-1)}\Ket{0}
\nonumber \\
&+&c_2\Ket{0}^{\otimes(L-1)}\Ket{0}\Ket{0}^{\otimes(L-1)}\Ket{1} \nonumber\\
&+&c_3\Ket{0}^{\otimes(L-1)}\Ket{1}\Ket{0}^{\otimes(L-1)}\Ket{0}\nonumber \\
&+&c_4\Ket{0}^{\otimes(L-1)}\Ket{1}\Ket{0}^{\otimes(L-1)}\Ket{1}.
\end{eqnarray}
After applying ${U^{\rm GHZ}}^{\otimes 2}$ on state $\Ket{\psi}$, one obtains
\begin{eqnarray}
\label{eq:appUghzstate}
{U^{\rm GHZ}}\otimes{U^{\rm GHZ}}\Ket{\psi}&=&c_1 \Ket{0}^{\otimes L}\Ket{0}^{\otimes L}
+c_2 \Ket{0}^{\otimes L}\Ket{1}^{\otimes L}\nonumber \\
&+&c_3 \Ket{1}^{\otimes L}\Ket{0}^{\otimes L}
+c_4 \Ket{1}^{\otimes L}\Ket{1}^{\otimes L}.\hphantom{aaaa}
\end{eqnarray}
The aim is to prepare a $2L^2$-qubit state, so one must introduce another 
$2L(L-1)$ ancillary qubits all initialized in the state $\Ket{0}$. The state 
Eq.~(\ref{eq:appUghzstate}) then becomes
\begin{eqnarray}
\Ket{\psi^\prime}&=&c_1 \left(\Ket{0}^{\otimes(L-1)}\Ket{0}\right)^{\otimes L} \left(\Ket{0}^{\otimes(L-1)}\Ket{0}\right)^{\otimes L} \nonumber \\
&+&c_2\left(\Ket{0}^{\otimes(L-1)}\Ket{0}\right)^{\otimes L} \left(\Ket{0}^{\otimes(L-1)}\Ket{1}\right)^{\otimes L} \nonumber\\
&+&c_3\left(\Ket{0}^{\otimes(L-1)}\Ket{1}\right)^{\otimes L} \left(\Ket{0}^{\otimes(L-1)}\Ket{0}\right)^{\otimes L} \nonumber \\
&+&c_4\left(\Ket{0}^{\otimes(L-1)}\Ket{1}\right)^{\otimes L} \left(\Ket{0}^{\otimes(L-1)}\Ket{1}\right)^{\otimes L}.
\end{eqnarray}
From Eqs.~(\ref{eq:HU*H00})(\ref{eq:HU*H01}), it is easy to see that
\begin{eqnarray}
\label{eq:UstarH0=G*}
&&U^{\rm star}H^{\otimes L} \Ket{0}^{\otimes L-1}\Ket{0}\\
&=&\left(\Ket{+}^{\otimes L-1}\Ket{0}+\Ket{-}^{\otimes L-1}\Ket{1}\right)
/{\sqrt{2}};\nonumber\\
&&U^{\rm star}H^{\otimes L} \Ket{0}^{\otimes L-1}\Ket{1}\\
&=&\left(\Ket{+}^{\otimes L-1}\Ket{0}-\Ket{-}^{\otimes L-1}\Ket{1}\right)/{\sqrt{2}}\nonumber.
\end{eqnarray}
State Eq.~(\ref{eq:UstarH0=G*}) is the $L$-qubit star graph state.

\begin{figure}[t]
    \centering
    \includegraphics[width=0.8\linewidth]{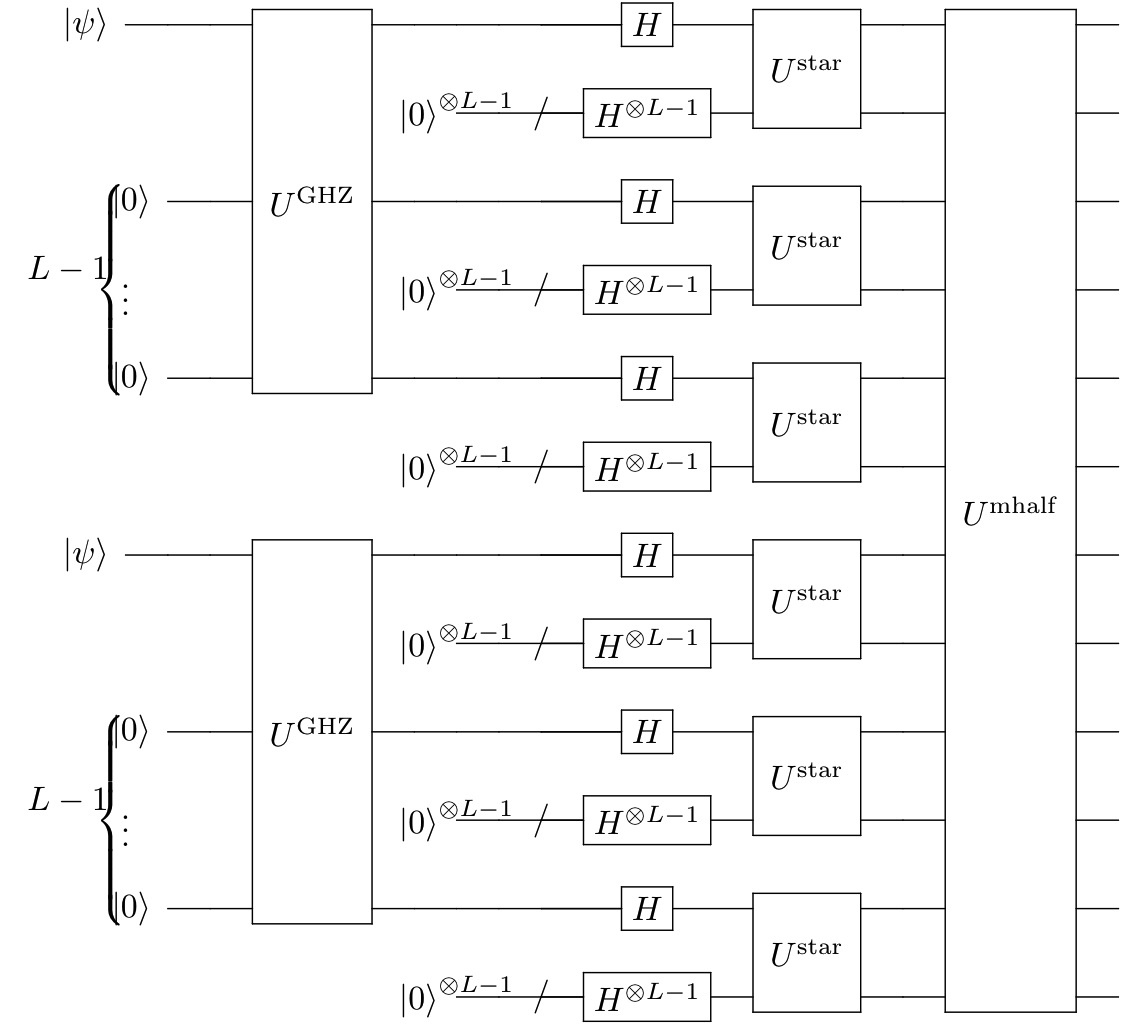}
    \caption{Quantum circuit encoding an arbitrary 2-qubit unknown state using code $C_{\rm gtoric}$ }
    \label{fig:qcircuit_encoding}
\end{figure}

Then we have
\begin{eqnarray}
\left(U^{\rm star}H^{\otimes L}\right)^{\otimes 2L}\Ket{\psi^\prime}&=&c_1 \Ket{\varphi^+_\star}\Ket{\varphi^+_\star}
+c_2 \Ket{\varphi^+_\star}\Ket{\varphi^-_\star}\nonumber \\
&+&c_3 \Ket{\varphi^-_\star}\Ket{\varphi^+_\star}
+c_4 \Ket{\varphi^-_\star}\Ket{\varphi^-_\star},\hphantom{aaaa}
\label{eq:encode_mstar_code}
\end{eqnarray}
where the states defined by
\begin{eqnarray}
\Ket{\varphi^+_\star} &=&\left(\frac{\Ket{+}^{\otimes L-1}\Ket{0}
+\Ket{-}^{\otimes L-1}\Ket{1}}{\sqrt{2}}\right)^{\otimes L},\\
\Ket{\varphi^-_\star} &=& \left(\frac{\Ket{+}^{\otimes L-1}\Ket{0}
-\Ket{-}^{\otimes L-1}\Ket{1}}{\sqrt{2}}\right)^{\otimes L}
\end{eqnarray}
satisfy 
\begin{eqnarray}
Z_{\alpha}\Ket{\varphi^+_\star}\otimes\Ket{\varphi^+_\star}
&=&\Ket{\varphi^-_\star}\otimes\Ket{\varphi^+_\star},\nonumber \\
Z_{\beta}\Ket{\varphi^+_\star}\otimes\Ket{\varphi^+_\star}
&=&\Ket{\varphi^+_\star}\otimes\Ket{\varphi^-_\star},\nonumber \\
Z_{\alpha}Z_{\beta}\Ket{\varphi^+_\star}\otimes\Ket{\varphi^+_\star}
&=&\Ket{\varphi^-_\star}\otimes\Ket{\varphi^-_\star};
\end{eqnarray}
here $Z_{\alpha}$ and $Z_{\beta}$ act on the last qubits of the first and 
second $L$ copies of star graph states, respectively.
Last, one must apply the multiple half-graph operator $U^{\rm mhalf}$ given in 
Eq.~(\ref{eq: U-mhalf}), which commutes with the 
$Z_\alpha$ and $Z_\beta$ because these operators only change the phase, as 
discussed in Sec.~\ref{subsec:half graph}. 
The whole quantum circuit is depicted in Fig.~\ref{fig:qcircuit_encoding}.
From the analysis in Sec.~\ref{sec:generation_toric_graph_state}, one has
$U^{\rm mhalf}\Ket{\varphi^+_\star}\Ket{\varphi^+_\star}=\Ket{G_{\rm toric}}$. 
The encoded logical state is therefore
\begin{eqnarray}
\label{eq:Umhalf-Gmstar}
\Ket{\psi_{\rm logical}}&=&U^{\rm mhalf}\left(
c_1 \Ket{\varphi^+_\star}\Ket{\varphi^+_\star}
+c_2 \Ket{\varphi^+_\star}\Ket{\varphi^-_\star}\right.\nonumber \\
&+&c_3 \Ket{\varphi^-_\star}\Ket{\varphi^+_\star}
+\left. c_4 \Ket{\varphi^-_\star}\Ket{\varphi^-_\star} \right) \nonumber\\
&=& \left(c_1+c_2 Z_\beta+c_3Z_\alpha+c_4 Z_\alpha Z_\beta \right)\Ket{G_{\rm toric}} \nonumber.
\end{eqnarray}
The quantum circuit for the full encoding procedure is depicted in 
Fig.~\ref{fig:qcircuit_encoding}. 

The procedure discussed above encodes an arbitrary two-qubit state in the code 
space spanned by the vectors
\begin{equation}
\operatorname{span}_{\mathbb{C}}\{\Ket{G_{\rm toric}}, 
Z_\alpha \Ket{G_{\rm toric}}, Z_\beta \Ket{G_{\rm toric}}, 
Z_\alpha Z_\beta \Ket{G_{\rm toric}} \},
\end{equation}
which is locally equivalent to toric code.
Without implementing the portion of the circuit corresponding to 
$U^{\rm mhalf}$ in the last step, the resulting state is instead given by
Eq.~(\ref{eq:encode_mstar_code}), which is the logical state of 
(\ref{eq:unknown_2qubit_state}) in the code space spanned by the vectors
\begin{equation}
\operatorname{span}_\mathbb{C}\{\Ket{\varphi^+_\star,\varphi^+_\star},
\Ket{\varphi^+_\star,\varphi^-_\star},
\Ket{\varphi^-_\star,\varphi^+_\star},\Ket{\varphi^-_\star,\varphi^-_\star}\}.
\end{equation}
This is a repetition code, as discussed in Sec.~\ref{sec:obser_toric_graph}.
Thus, the repetition encoding procedure is a subroutine of full encoding 
circuit for the toric code.

\subsection{State preparation via measurements}

Any graph can be expanded to a graph containing vertices with at most degree
three, by introducing ancillae~\cite{hoyer2006} which are then measured in the
$X$ basis. Given that all the measurements commute and can therefore be
performed simultaneously, graph state preparation can be performed in 
(constant) depth three. For example, the star graphs on $L$ qubits with degree 
$L-1$ that are induced subgraphs of the toric graph can be represented by a 
totally asymmetric tree graph with $2L-4$ vertices with maximum degree three. 
Given that there are $2L$ star graphs comprising the toric graph, the 
contribution of the star-graph ancillae to the circuit width scales as 
$L^2\sim N$. 

Next consider the half graphs on $2(L-1)$ vertices that constitute the 
remaining subgraphs. The first vertex in the first of the two bipartite vertex 
subsets has an edge with all $L-1$ vertices in the second vertex subset and 
thus has the connectivity of a degree-$L-1$ star graph; following the procedure 
described for the star graphs, one can add $L-4$ ancillae to this central 
vertex to ensure that all resulting vertices have at most degree three. The 
second vertex in the first subset shares $L-2$ neighbors in the second subset,
requiring the addition of $L-5$ ancillae, etc. The total number of required
ancillae therefore scales as $L^2$; and, given that the number of half graphs
scales as $L$, the total number of half-graph ancillae scales as $L^3$. Thus,
preparing the toric graph of size $N$ in constant depth as a maximum 
degree-three graph requires a circuit width 
${\mathcal O}(L^3)\sim{\mathcal O}(N^{3/2})$.

It is worthwhile to point out that the ${\mathcal O}(L^3)$-qubit graph state 
with bounded degree-three vertices is topologically trivial, while the state 
that remains after the $X$ measurements is the topologically ordered toric 
graph state. Thus, one can obtain a topologically ordered state by projective 
measurement on a topologically trivial state in a higher-dimensional Hilbert 
space.

There are other methods to prepare one of the logical states in the toric code 
state by measurement. For example, one can perform projective measurements for 
all star operators and $S_\beta$ on the state $\Ket{0}^{\otimes N}$, then 
correct the signs according to the measurement result. There is also a 
measurement-based method to generate toric code 
states~\cite{Zarei2017,Bolt2016}, but the depth is not analyzed explicitly.

\section{Discussion}
\label{sec:discussion}

In this work, we map a toric code state to its LC-equivalent graph state, which 
is found to consist solely of star and half graphs. Given that the star graphs 
encode GHZ states, the graph construction reveals a 
novel connection between the toric code and the nine-qubit (repetition) error 
correcting code, which is itself an instance of a family of codes with 
macroscopic code distance. The star graphs therefore contribute the large code 
distance, while the half graphs ensure that the toric code stabilizer 
generators are low-weight and geometrically local. It was shown in 
Ref.~\cite{Bombin2012Universal} not only that all 2D translationally invariant 
topological stabilizer codes are in the same universal phase, but also that 
they can all be mapped to multiple copies of the toric code. This implies that
the graph states that are LC-equivalent to 2D topological stabilizer states are 
also characterized by star graphs and half graphs.

We also construct an explicit quantum circuit which generates the toric 
code states on $N$ qubits in $\log(N)$ depth, or in constant depth if the 
circuit width is allowed to increase as $N^{3/2}$, under the assumption of 
geometrically non-local gates. By using the local-Clifford
equivalence of stabilizer and graph states, the quantum circuit is obtained as 
an efficient set of gates that effects the quadratic Boolean function 
associated with the graph adjacency matrix. These results can also be used to 
generate any 2D topological stabilizer state with translation symmetry, based 
on the results in Ref.~\cite{Bombin2012Universal}.

A fruitful avenue for future research would be to probe what features of the 
toric graph are specific to the toric code, and what features (if any) are 
necessary to ensure that the graph connectivity encodes a state with 
non-trivial topological order. For example, the multiple copies of the star
graph are sufficient to ensure that the code distance increases polynomially
with the number of physical qubits. Presumably other subgraphs could accomplish 
the same, but potentially at the cost of increasing the minimum weight of the 
associated stabilizer generators. In any case, constructing the graph state 
first and then reversing the mapping to the stabilizer set could allow for the
discovery of new classes of topological error-correction codes.

\acknowledgments
The authors are grateful to Barry Sanders for insightful comments. This work 
was supported by the Natural Sciences and Engineering Research Council of 
Canada. 

\appendix

\section{Multiple copies of the GHZ state}
\label{sec:proof}

\begin{remark}
Given two $m$-qubit GHZ states
\begin{align}
\Ket{\phi_m^+} = \frac{\Ket{0^m}+\Ket{1^m}}{\sqrt{2}},\quad
\Ket{\phi_m^-} = \frac{\Ket{0^m}-\Ket{1^m}}{\sqrt{2}}.
\end{align}
and their $m$-copy states on $m^2$ qubits
\begin{equation}
\label{eq:m_copy_GHZ_appen}
\Ket{\varphi^+} =  \Ket{\phi_m^+}^{\otimes m},
\Ket{\varphi^-} =  \Ket{\phi_m^-}^{\otimes m}.
\end{equation}
Then, $\operatorname{span}_{\mathbb{C}}\Set{\Ket{\varphi^+},
\Ket{\varphi^-}}$ is a quantum error correction code with distance $d=m$.
\end{remark}

\begin{proof}
The weight of the operator $O\in\mathcal{P}_N$ is the number of qubits which 
are acted on non-trivially (i.e.\ by a non-identity) by $O$. Here,
$\mathcal{P}_N^d\subset \mathcal{P}_N$ contains all the operators whose weight 
is less than $d$. Based on the quantum error correction 
condition~\cite{Knill1997,nielsen2000}, the subspace 
$\operatorname{span}_{\mathbb{C}}\set{\Ket{\psi_1},\Ket{\psi_2}} \subset 
\mathcal{H}_2^{\otimes N}$ is a quantum error correction code with distance $d$ 
if and only if the following conditions always hold:
\begin{align}
\label{eq:qerror_condition_1}
\braket{\psi_1|O|\psi_1} =& \braket{\psi_2|O|\psi_2},  \\
\label{eq:qerror_condition_2}
\braket{\psi_1|O|\psi_2} =& 0,
\end{align}
for every operator $O\in\mathcal{P}_N^d$.

It is straightforward to verify that 
$\braket{\phi_m^+|O|\phi_m^+}=\braket{\phi_m^-|O|\phi_m^-}$ is satisfied 
$\forall O\in\mathcal{P}_m^m$, while the second condition fails to hold because 
$\braket{\phi_m^+|Z_i|\phi_m^-}=1$, where $Z_i$ is the Pauli $Z$ operator 
acting on qubit $i$.  Consider multiple copies of the GHZ state instead. One
can again verify that $\braket{\varphi^+|O|\varphi^+}
=\braket{\varphi^-|O|\varphi^-}$ still holds $\forall O\in\mathcal{P}_{m^2}^m$.
On the other hand, one obtains
\begin{align}
\braket{\varphi^+|O|\varphi^-} = 
\braket{\varphi^+|\bigotimes_{i=1}^m O_i|\varphi^-} = 
\prod_{i=1}^m \braket{\phi_m^+|O_i|\phi_m^-}, 
\end{align}
where $O_i\in\mathcal{P}_m$ acts on the $i$-th copy. If
$\braket{\varphi^+|O|\varphi^-}\ne0$, then 
$\braket{\phi_m^+|O_i|\phi_m^-}\ne0,\forall i\in\set{1,\ldots,m}$,
so $O$ acts on at least $m$ qubits non-trivially and  $O\notin P_{m^2}^m$.
Therefore, condition Eq.~(\ref{eq:qerror_condition_1}) and Eq.~(\ref{eq:qerror_condition_2}) hold $\forall O\in\mathcal{P}_{m^2}^m$ for $\Ket{\varphi^+}$ and $\Ket{\varphi^-}$.
\end{proof}

\section{Quantum circuit generating star graph states}
\label{sec:appen_algirothm_state_graph_state}

In this section, we prove that Algorithm~\ref{alg:generate_star_graph_state} 
generates a star graph state in log depth. Line 1 to line 5 describe the 
initialization process and the state in quantum register after line 5 is 
$H^{\otimes n}\Ket{0^n}$. Next, the circuit described from 
line~\ref{line:start_CX_loop} to line~\ref{line:end_CX_loop} is the gate $P$ in 
Eq.~(\ref{eq:U_f_s}) that implement $\operatorname{Parity}(q_1,\ldots,q_{n-1})$.
Line~\ref{line:CZ} adds a global phase 
$(-1)^{\operatorname{Parity}(q_1,\ldots,q_{n-1})\cdot q_n}$ and the remaining 
operation is exactly inverse of $P$. 
It only remains to prove that line~\ref{line:start_CX_loop} to 
line~\ref{line:end_CX_loop} indeed implements 
$\operatorname{Parity}(q_1,\ldots,q_{n-1})$.

\begin{widetext}
\begin{figure*}[t]
\begin{minipage}{\linewidth}
\begin{algorithm}[H]
\begin{algorithmic}[1]
\Require
{
\Statex \texttt{unsignedinteger} \textsc{num} \Comment{number of qubits}
}
\Ensure 
{ 
\Statex  \underline{\texttt{binary}}[\textsc{num}] \underline{\textsc{qsTate}}  \Comment{ \textsc{num}-qubit star graph state}
}
\Function {Generate Star Graph State}{\textsc{num}}
\State \texttt{unsignedinteger} \textsc{dePth}
\Comment{Depth of the circuit}
\State \texttt{unsignedinteger} \textsc{qtArg}
\Comment{Index of target qubit}
\State \underline{\textsc{qsTate}} $\gets \underline{0}^{\frown \textsc{num}}$
\Comment{Initialize quantum register}
\State \underline{\textsc{qsTate}} $\gets \operatorname{H}[\textsc{num}]$ \underline{*}\underline{\textsc{qsTate}}
\Comment{Apply Hadamard gate}
\State \textsc{dePth}  $\gets\lceil \log(\textsc{num}-1)\rceil $ \label{line:start_CX_loop}
\For{$d$ from 1 to in \textsc{dePth} } 
\For{$i$ from $2^{d-1}$ to \textsc{num}-1 }
\If{$i=2^{d-1}  \mod 2^d$}
\State  \textsc{qtArg} $\gets \min \{i+2^{d-1},\textsc{num}-1\}$ 
\label{line:target_qubit}
\State  \underline{\textsc{qsTate}} $\gets \operatorname{CX}(i, \textsc{qtArg} )$ \underline{*}  \underline{\textsc{qsTate}}
\Comment{Apply CX gate}
\EndIf
\EndFor
\EndFor \label{line:end_CX_loop}
\State \underline{\textsc{qsTate}} $\gets$ {CZ}$(\textsc{num}-1,\textsc{num})$ \underline{*}  \underline{\textsc{qsTate}}
\label{line:CZ}
\Comment{Apply CZ gate}
\For{$d$ from  \textsc{dePth} to 1} \label{line:start_uncompute}
\For{$i$ from $2^{d-1}$ to \textsc{num} }
\If{$i-2^{d-1} = 0$  mod  $2^d$}
\State  \textsc{qtArg} $\gets \min \{i+2^{d-1},\textsc{num}-1\}$ 
\State  \underline{\textsc{qsTate}} $\gets$ {CX}$(i,\textsc{qtArg})$ \underline{\textsc{qsTate}}
\EndIf
\EndFor
\EndFor \label{line:end_uncompute}
\EndFunction
\end{algorithmic}
\caption{Generate star graph state} 
\label{alg:generate_star_graph_state} 
\end{algorithm}
\end{minipage}
\end{figure*}

\begin{figure*}[t]
\begin{minipage}{\linewidth}
\begin{algorithm}[H] 
\caption{Generate half graph state} 
\begin{algorithmic}[1]
\Require
{
\Statex \texttt{unsignedinteger} \textsc{num} \Comment{size of the half graph}
}
\Ensure 
{ 
\Statex  \underline{\texttt{binary}}[2\textsc{num}] \underline{\textsc{qsTate}}  
\Comment{ 2\textsc{num}-qubit half graph state}
}
\Function {Generate half}{\textsc{num}}
\State \texttt{unsignedinteger} \textsc{dePth}
\Comment{Depth of the circuit}
\State \texttt{unsignedinteger} \textsc{coNxqBit}
\State \texttt{unsignedinteger} \textsc{coNyqBit}
\State \texttt{unsignedinteger} \textsc{taRxqBit}
\State \texttt{unsignedinteger} \textsc{taRyqBit}
\State \underline{\textsc{qsTatEx}} $\gets \underline{0}^{\frown \textsc{num}}$
\Comment{Quantum register for $\Ket{q}$}
\State  \underline{\textsc{qsTatEx}} $\gets$ \underline{\textsc{Had}}$(\textsc{num})$ \underline{*}  \underline{\textsc{qsTatEx}}
\State \underline{\textsc{qsTatEy}} $\gets \underline{0}^{\frown \textsc{num}}$
\Comment{Quantum register for $\Ket{p}$}
\State  \underline{\textsc{qsTatEy}} $\gets$ \underline{\textsc{Had}}$(\textsc{num})$ \underline{*}  \underline{\textsc{qsTatEy}}
\For{$i$ from 1 to $\textsc{num}$ } \label{line:start_f^0}
\State \textsc{coNxqBit} $\gets i$
\State \textsc{taRyqBit} $\gets i$
\State \underline{\textsc{qsTate}}$\gets$ \underline{CZ}$(\textsc{coNxqBit},\textsc{taRyqBit})$ \underline{*}  (\underline{\textsc{qsTatEx}}, \underline{\textsc{qsTatEy}})
\EndFor \label{line:end_f^0}

\State \textsc{dePth}  $\gets\lceil \log(\textsc{num})\rceil $ \label{line:half_start_iteration}
\For{$d$ from 1 to in \textsc{dePth} }
\For{$i$ from $1$ to $\textsc{num}$ }
\If{$i=2^{d-1}\mod2^d$}
\State \textsc{coNxqBit} $\gets i$
\State \textsc{coNyqBit} $\gets i$
\State \textsc{taRxqBit} $\gets \min\{i+ 2^{d-1},\textsc{num}\}$ \label{line:upper_bound_x}
\State \textsc{taRyqBit} $\gets \min\{i+ 2^{d-1},\textsc{num}\}$ \label{line:upper_bound_y}

\State \underline{\textsc{qsTate}}$\gets$ \underline{CZ}$(\textsc{coNxqBit},\textsc{taRyqBit})$ \underline{*}  (\underline{\textsc{qsTatEx}}, \underline{\textsc{qsTatEy}})
\State  \underline{\textsc{qsTatEx}} $\gets$ \underline{CX}$(\textsc{coNxqBit},\textsc{taRxqBit})$ \underline{*}  \underline{\textsc{qsTatEx}}
\State  \underline{\textsc{qsTatEy}} $\gets$ \underline{CX}$(\textsc{coNyqBit},\textsc{taRyqBit})$ \underline{*}  \underline{\textsc{qsTatEy}}
\EndIf
\EndFor
\EndFor \label{line:half_end_iteration}
\For{$d$ from  \textsc{dePth} to 1  } 
\label{line:start_half_uncomput}
\For{$i$ from $2^{d-1}$ to $\textsc{num}$ }
\If{$i-2^{d-1} = 0$  mod  $2^d$}
\State \textsc{coNxqBit} $\gets i$
\State \textsc{coNyqBit} $\gets i$
\State \textsc{taRxqBit} $\gets \min\{i+ 2^{d-1},\textsc{num}\}$
\State \textsc{taRyqBit} $\gets \min\{i+ 2^{d-1},\textsc{num}\}$
\State  \underline{\textsc{qsTatEx}} $\gets$ \underline{CX}$(\textsc{coNxqBit},\textsc{taRxqBit})$ \underline{*}  \underline{\textsc{qsTatEx}}
\State  \underline{\textsc{qsTatEy}} $\gets$ \underline{CX}$(\textsc{coNyqBit},\textsc{taRyqBit})$ \underline{*}  \underline{\textsc{qsTatEy}}
\EndIf
\EndFor
\EndFor \label{line:end_half_uncomput}
\EndFunction
\end{algorithmic}
\label{alg:generat_half_state} 
\end{algorithm}
\end{minipage}
\end{figure*}
\end{widetext}

Consider the action from line~\ref{line:start_CX_loop} to line~\ref{line:end_CX_loop} on input $\Ket{q}$, a computational basis state and  $q\in\set{0,1}^n$. 
$d$ denotes the times of iteration and $\Ket{q^d}$ is the state in quantum register after the $d$-th iteration.
During the $d$-th iteration, the gate $\operatorname{CX}(2^{d-1}+c2^d,(c+1)2^{d})$, where $c\in\mathbb{N}$, is executed.
One notices that all $\operatorname{CX}$ gates within the same iteration act on no common qubits so they can all be performed in parallel.
The depth of the circuit from line~\ref{line:start_CX_loop} to line~\ref{line:end_CX_loop} is then $\log(n-1)$.

After $d$-th iteration, the state in the quantum register should satisfy $q_{c2^d}^d =\sum^{c2^d}_{j=(c-1)2^d+1} q_{j}$, where $c\in\{1,\ldots,2^{\lceil \log(n-1)\rceil-d}\}$.
We prove this claim using induction. First, it holds trivially when $d=0$. 
Next, suppose it holds when $d=k$.
Then in the $(k+1)$-th loop, gate $\operatorname{CX}(c2^{k+1}-2^k,c2^{k+1})$ is executed, where $c\in\{1,\ldots,2^{\lceil \log(n-1)\rceil-{k-1}}\}$, so
\begin{equation}
\begin{aligned}
&q_{c2^{k+1}}^{k+1} =  q_{c2^{k+1}}^{k}+q_{c2^{k+1}-2^k}^k 
=  q_{2c2^{k}}^{k}+q_{(2c-1)2^{k}}^k \\
=& \sum^{2c2^k}_{j=(2c-1)2^k+1} q_{j} + \sum^{(2c-1)2^k}_{j=(2c-2)2^k+1} q_{j}\\
=& \sum^{c2^{k+1}}_{j=(c-1)2^{k+1}+1} q_{j}.
\end{aligned}
\end{equation}

If $\log_2(n-1)\in\mathbb{N}$, then after $\log_2(n-1)$-th iteration, $q_{n-1}^{\log_2(n-1)}=\operatorname{Parity}(q_1,\ldots,q_{n-1})$. 
When $\log_2(n-1)\notin\mathbb{N}$, then in the $\log_2(n-1)$-th iteration, the target qubit is replace as $n-1$-th qubit when it exceed $n-1$ in line~\ref{line:target_qubit}, so $q_{n-1}^{\lceil\log_2(n-1) \rceil}=\operatorname{Parity}(q_1,\ldots,q_{n-1})$ still holds. 
Therefore, we conclude Algorithm~\ref{alg:generate_star_graph_state} indeed generates a star graph state in log depth.

\section{Quantum circuit generating  half states}
\label{sec:appendix_qcircuit_half_graph_state}

The Unitary described from line~\ref{line:start_f^0} to line~\ref{line:end_f^0} adds the phase $(-1)^{f_h^0(q,p)}$.
Next, consider what unitary lines~\ref{line:half_start_iteration}-\ref{line:half_end_iteration} computes when the input is in the computational basis $\Ket{q,p}$. 
$\Ket{q^{d},p^{d}}$ denotes the state in the register after the $d$-th iteration. 
The operation within the quantum register for $\Ket{q}$ ( $\Ket{p}$) is the same as in Algorithm 1, so after $k-1$-th iteration
\begin{align}
q_{c2^{k-1}}^{k-1}= \sum_{j=(c-1)2^{k-1}+1}^{c2^{k-1}} q_{j},\\
p_{c2^{k-1}}^{k-1} = \sum_{j=(c-1)2^{k-1}+1}^{c2^{k-1}} p_{j}, 
\end{align}
holds.

In addition, in the $k$-th iteration, gate $\operatorname{CZ}(q_{c2^k-2^{k-1}},p_{c2^k})$ is executed with the quantum register in state $\Ket{q^{k-1},p^{k-1}}$,
adding a phase $(-1)^{f_k^\prime(q,p)}$, where
\begin{equation}
\begin{aligned}
&f_k^\prime(q,p) = \sum_{c=1}^{n/2^k} q_{c2^k-2^{k-1}}^{k-1}\cdot p_{c2^k}^{k-1}\\
&= \sum_{c=1}^{n/2^k} \left(\sum_{j=(2c-2)2^{k-1}+1}^{(2c-1)2^{k-1}} q_{j}\right)\cdot \left(\sum_{j=(2c-1)2^{k-1}+1}^{2c2^{k-1}} p_{j}\right)\\
&= \sum_{c=0}^{n/2^k-1} \left(\sum_{j=c2^{k}+1}^{(c+1/2)2^{k}} q_{j}\right)\cdot \left(\sum_{j=(c+1/2)2^{k}+1}^{(c+1)2^{k}} p_{j}\right)\\
&= f^k_h(q,p).
\end{aligned}
\end{equation}
When $n$ is not the power of two, the sum index upper bound is replaced in line~\ref{line:upper_bound_x}  and line~\ref{line:upper_bound_y}.
Therefore, the quantum circuit in the $k$-th iteration adds a global phase $(-1)^{f^k_h(q,p)}$ and 
all $\operatorname{CZ}$ ($\operatorname{CX}$) gates within the same iteration can be performed in parallel.
Lines~\ref{line:start_half_uncomput}-\ref{line:end_half_uncomput} uncompute the garbage and leave only the phase. 
Because of the decomposition Eq.~(\ref{eq:decomposition_half_boolean_function}), this algorithm prepares a half graph state in log depth.

\bibliographystyle{apsrev.bst}
\bibliography{ref}{}
\end{document}